\documentclass[12pt]{article}
\usepackage{graphicx,bm,amsmath,amssymb,latexsym,epsfig,euscript,multirow,color,url}
\usepackage{hyperref}
\allowdisplaybreaks

\long\def\symbolfootnote[#1]#2{\begingroup%
\def\thefootnote{\fnsymbol{footnote}}\footnote[#1]{#2}\endgroup}

\newcommand{\be}{\begin{equation}}
\newcommand{\ee}{\end{equation}}
\newcommand{\pd}{\partial}
\newcommand{\bea}{\begin{eqnarray}}
\newcommand{\eea}{\end{eqnarray}}
\newcommand{\mat}{\begin{pmatrix}}
\newcommand{\rix}{\end{pmatrix}}
\newcommand{\nn}{\nonumber}
\renewcommand{\l}{\left}
\renewcommand{\r}{\right}

\renewcommand{\bar}{\overline}
\renewcommand{\slash}[1]{#1\!\!\!/}

\newcommand{\vv}{\mathbf}

\newcommand{\cM}{\mathcal{M}}
\newcommand{\cL}{\mathcal{L}}

\newcommand{\gr}{{\tilde G}}

\newcommand{\st}{{\tilde t}}

\newcommand{\q}{\quad}
\newcommand{\qq}{\qquad}

\newcommand{\comment}[1]{}

\input prepictex
\input pictex
\input postpictex
\newdimen\tdim
\tdim=\unitlength
\def\stpltsmbl{\setplotsymbol ({\small .})}

\begin{document}

\begin{titlepage}

\begin{flushright}
\small{RUNHETC-2011-12}\\
\end{flushright}

\vspace{0.5cm}
\begin{center}
\Large\bf
Light Stop NLSPs at the Tevatron and LHC
\end{center}

\vspace{0.2cm}
\begin{center}
{\sc Yevgeny Kats\symbolfootnote[1]{kats@physics.rutgers.edu} and David Shih\symbolfootnote[2]{dshih@physics.rutgers.edu}}\\
\vspace{0.6cm}
\textit{New High Energy Theory Center\\
Department of Physics and Astronomy\\
Rutgers University, Piscataway, NJ 08854, USA}
\end{center}

\vspace{0.5cm}
\begin{abstract}
\vspace{0.2cm}
\noindent

How light can the stop be given current experimental constraints? Can it still be lighter than the top? In this paper, we study this and related questions in the context of gauge-mediated supersymmetry breaking, where a stop NLSP decays into a $W$, $b$ and gravitino. Focusing on the case of prompt decays, we simulate several existing Tevatron and LHC analyses that would be sensitive to this scenario, and find that they allow the stop to be as light as $150$~GeV, mostly due to the large top production background. With more data, the existing LHC analyses will be able to push the limit up to at least $180$~GeV. We hope this work will motivate more dedicated experimental searches for this simple scenario, in which, for most purposes, the only free parameters are the stop mass and lifetime.

\end{abstract}
\vfil

\end{titlepage}

\tableofcontents

\section{Introduction}
\setcounter{equation}{0}

In supersymmetric extensions of the Standard Model, there are typically many reasons to suppose that the stop is the lightest squark. These reasons include: electroweak-scale baryogenesis in the MSSM (for a recent analysis, see~\cite{Carena:2008vj}); the little hierarchy problem (see, e.g.,~\cite{Kitano:2006gv,Asano:2010ut} for a recent discussion of this); and the fact that light stops arise naturally in the MSSM -- renormalization-group running and level-splitting generally pushes the third generation lighter than the first and second generations (see~\cite{Martin:1997ns} for more details). Finally, probably the most important motivation of all in the LHC-discovery era  -- light third generations can lead to very interesting and less explored collider signatures.

Light stops in theories of gauge-mediated supersymmetry breaking (GMSB) are an especially interesting and motivated possibility. As is well known, gauge mediation is an appealing supersymmetric scenario: it automatically solves the flavor problem, and it generates phenomenologically viable soft masses. In such theories, the lightest superpartner (LSP) is always a nearly-massless gravitino $\gr$. Assuming $R$-parity, the next-to-lightest superpartner (NLSP) decays in a universal fashion to the gravitino plus its Standard Model partner. Recently, a model-independent framework for general gauge mediation (GGM) was established in~\cite{Meade:2008wd,Buican:2008ws}. In GGM, essentially any MSSM superpartner can be the NLSP. So it is interesting to consider the case that the NLSP is the lightest stop $\st$. The dominant decay of the stop in such a scenario is
\be
\st\,\,\, \to\,\,\, W^+ b  \gr
\label{stop-decay}
\ee
Intriguingly, despite the fact that this possibility has been known for more than a decade~\cite{Sarid:1999zx,Chou:1999zb,Demina:1999ty}, no searches have addressed it explicitly. And this scenario is far from being obviously excluded.

\begin{figure}[t]
\begin{center}
\includegraphics[width=0.43\textwidth]{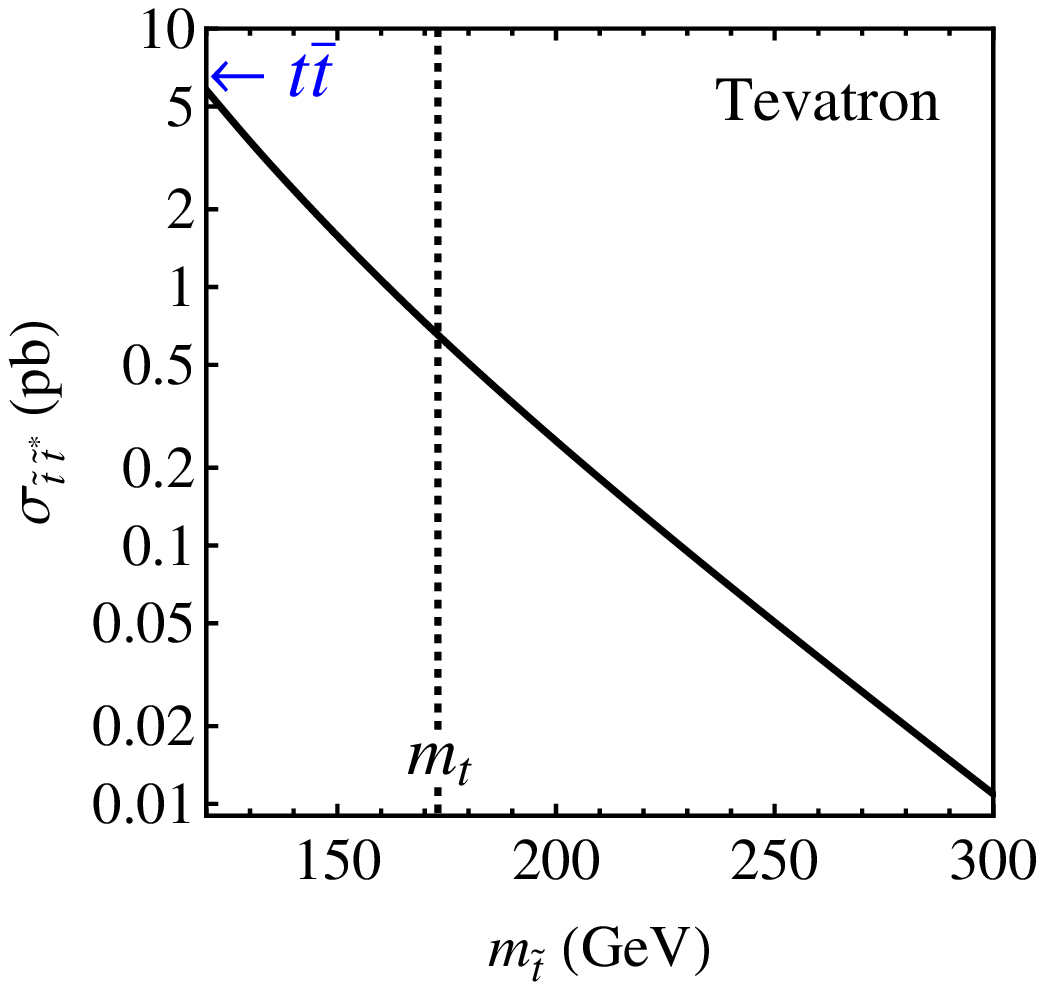}\q
\includegraphics[width=0.43\textwidth]{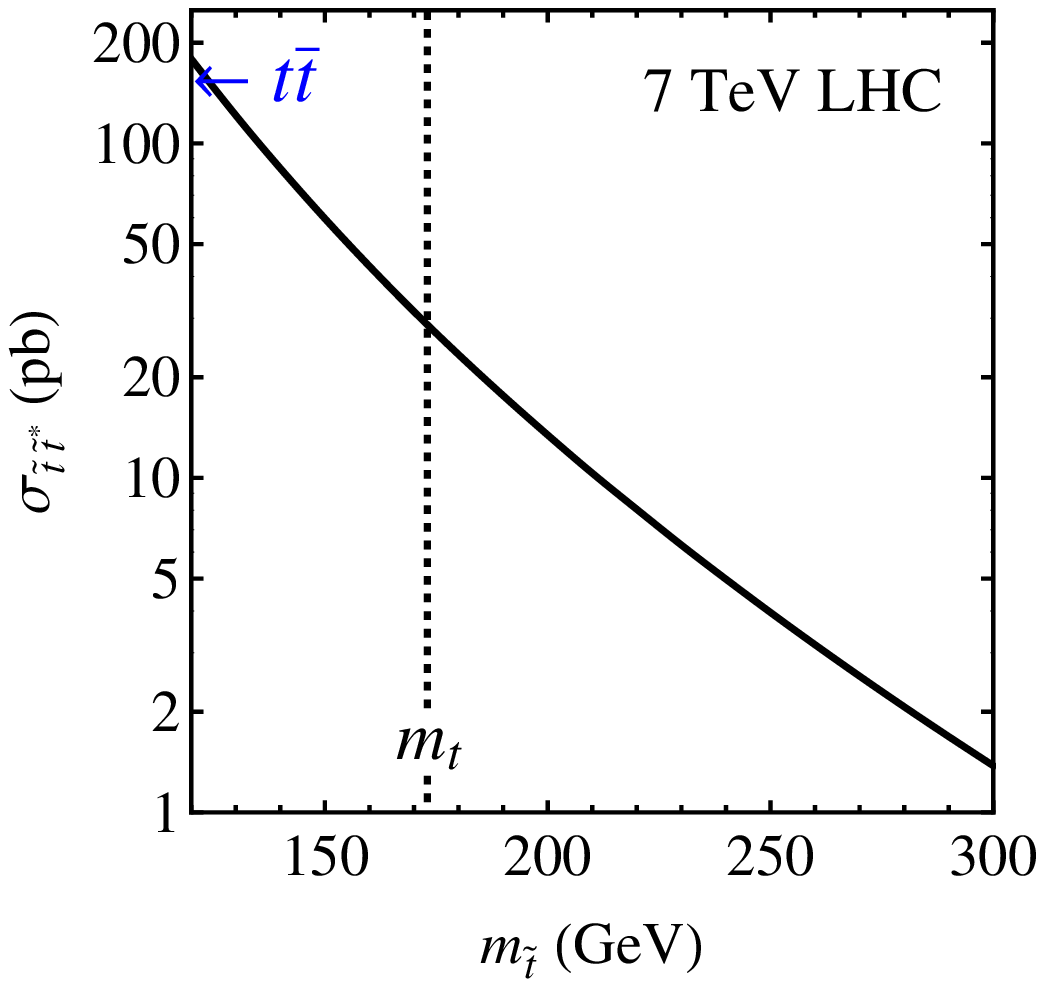}
\end{center}
\caption{The NLO+NLL stop pair production cross section at the Tevatron (left) and $7$~TeV LHC (right) as a function of the stop mass. The values of $t\bar t$ cross sections are indicated as well. For more details, see appendix~\ref{sec-xsec}.}
\label{fig-stop-cs}
\end{figure}

In this paper, we will focus on the following simple question: how light can the stop NLSP be without being in conflict with existing data? In particular, can the stop be lighter than the top? Since the stop is colored, stop-antistop pairs have sizeable production cross sections at hadron colliders, especially if the stop is light. Still, they can be missed if their decay products have a large Standard Model background. Indeed, $t\bar t$ production (where $t\to W^+b$) has a very similar signature to $\st\st^\ast$ production, with a much larger cross section  (see figure~\ref{fig-stop-cs}). Meanwhile the uncertainties on the top cross section, both experimental and theoretical, are of the order of $10\%$. As a result, the stop signal may not stand out  in $t\bar t$ cross section measurements that use simple cuts and event counting. On the other hand, more sophisticated measurements of the $t\bar t$ cross section and other properties of the top, especially those that require the detailed reconstruction of the $t\bar t$ event, may dismiss the stop events as background. One of the main goals of this paper is to re-analyze some of the existing Tevatron and LHC measurements of $t\bar t$-like samples and find out to what extent they constrain the stop NLSP scenario.

For simplicity, we will study a minimal spectrum, consisting solely of stop NLSP $\st$ and a nearly massless gravitino $\tilde G$. We will take all the other states of the MSSM to be decoupled, which is technically possible in the GGM parameter space. This results in two simplifications. First, we neglect the contribution from similar decays of the second stop mass eigenstate or the production of stops from the decays of other colored states (such as gluinos). Therefore, for scenarios in which these additional particles are relatively light, the stop NLSP limits derived here should be thought of as somewhat conservative. Second, the decoupling assumption fixes the diagrams contributing to the stop decay process~(\ref{stop-decay}) to those shown in figure~\ref{fig-diagrams}. This will not limit the generality of our conclusions since the basic kinematic properties of the stop events do not depend much on the relative contributions of the various diagrams~\cite{Chou:1999zb}.

We start in section~\ref{sec-stop-NLSP} by discussing the properties of the stop NLSP decay process. In section~\ref{sec-stop-searches} we discuss the existing Tevatron and LHC analyses that may be sensitive to this scenario and use several of them for deriving our constraints in section~\ref{sec-results}. Section~\ref{sec-other-measurements} discusses several other types of measurements that may be relevant to stop NLSPs. We conclude in section~\ref{sec-conclusions}.

\section{Phenomenology of stop NLSP\label{sec-stop-NLSP}}
\setcounter{equation}{0}

\subsection{The decay process}

\begin{figure}[t]
$$\beginpicture
\setcoordinatesystem units <0.6\tdim,0.6\tdim>
\stpltsmbl
\setdashes
\plot -60 0 0 0 /
\setdots
\plot 0 0 60 -50 /
\setsolid
\plot 0 0 96 48 /
\startrotation by 0.7 -0.55 about 40 20
\ellipticalarc axes ratio 2:1 -220 degrees from  40 20 center at  50 20
\ellipticalarc axes ratio 2:1 -280 degrees from  57 16 center at  64 20
\ellipticalarc axes ratio 2:1 -280 degrees from  72 16 center at  79 20
\ellipticalarc axes ratio 2:1 -280 degrees from  87 16 center at  94 20
\ellipticalarc axes ratio 2:1 -180 degrees from 102 16 center at 109 20
\stoprotation
\put {$\st$} at -75 5
\put {$b$} at 110 55
\put {$W^+$} at 120 -20
\put {$\gr$} at 78 -50
\put {$t^{(\ast)}$} at 18 25
\linethickness=0pt
\putrule from 0 -65 to 0 65
\putrule from -150 0 to 150 0
\endpicture
\beginpicture
\setcoordinatesystem units <0.55\tdim,0.55\tdim>
\stpltsmbl
\setdashes
\plot -60 0 0 0 /
\setdots
\plot 0 0 60 -50 /
\setsolid
\plot 0 0 60 50 /
\ellipticalarc axes ratio 2:1 -220 degrees from  0  0 center at 10 0
\ellipticalarc axes ratio 2:1 -280 degrees from 17 -4 center at 24 0
\ellipticalarc axes ratio 2:1 -280 degrees from 32 -4 center at 39 0
\ellipticalarc axes ratio 2:1 -280 degrees from 47 -4 center at 54 0
\ellipticalarc axes ratio 2:1 -180 degrees from 62 -4 center at 69 0
\put {$\st$} at -75 5
\put {$b$} at 75 55
\put {$W^+$} at 100 5
\put {$\gr$} at 78 -50
\linethickness=0pt
\putrule from 0 -65 to 0 65
\putrule from -150 0 to 150 0
\endpicture$$
\caption{Diagrams contributing to the stop NLSP decay in GMSB in the simplified scenario where all the other superpartners are heavy.}
\label{fig-diagrams}
\end{figure}

\begin{figure}[t]
\begin{center}
\includegraphics[width=0.418\textwidth]{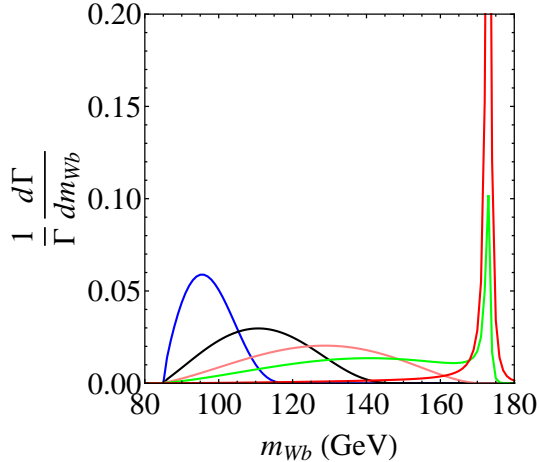}
\end{center}
\caption{Distributions of the $Wb$ invariant mass for stops with masses (in GeV) $120$ (blue), $150$ (black), $172$ (pink), $180$ (green) and $200$ (red). We assume $m_t = 173$~GeV.}
\label{fig-mWb}
\end{figure}

In the limit that all of the other superpartners besides the lightest stop are decoupled, only the diagrams shown in figure~\ref{fig-diagrams} contribute to the stop decay process~(\ref{stop-decay}). These two diagrams arise from the same gauge-covariant derivative, as explained in more detail in appendix~\ref{sec-ME}. The resulting matrix element is
\be
\cM = i\frac{\sqrt2}{F}\,g\,
\bar u(p_b)\l[\l(p_\gr\cdot \epsilon_W^\ast\r) c_\st P_R - \frac{1}{2}\frac{m_\st^2 - m_{Wb}^2}{m_{Wb}^2 - m_t^2 + im_t\Gamma_t}\,\slash \epsilon_W^\ast \l(m_\st c_\st \gamma^0 P_R + m_t s_\st P_L\r)\r] v(p_\gr)
\label{ME}
\ee
where $\sqrt F$ is the SUSY breaking scale, $g$ is the $SU(2)$ gauge coupling, and $c_\st$, $s_\st$ describe the mixing in the stop sector. For $m_\st\lesssim m_t$, the contributions of the two diagrams are comparable, while for higher masses the diagram involving the top starts to dominate and eventually reduces to the 2-body decay $\st\to t\gr$. The transition between the 3-body decay and the 2-body decay is demonstrated in figure~\ref{fig-mWb}.

From (\ref{ME}), we see that our model depends on just two parameters: the mass of the lighter stop and the stop mixing angle. For a more general spectrum, diagrams with virtual charginos or sbottoms would also contribute, but as was noticed in~\cite{Chou:1999zb}, where a much larger parameter space has been explored, the kinematic distributions, such as the invariant masses $m_{\ell b}$ and $m_{bW}$, do not depend strongly on the assumptions about the spectrum or the stop mixing angle (which we will set to $s_\st = -0.8$). We therefore believe that the simplified scenario we consider is a good representative of the whole class of stop NLSP scenarios.

\begin{figure}[t]
\begin{center}
\includegraphics[width=0.55\textwidth]{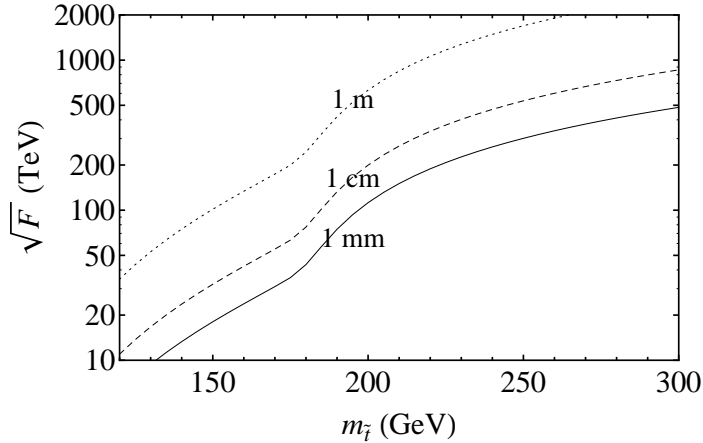}
\end{center}
\caption{Contours of constant $c\tau$ for stop NLSP decay, as a function of the SUSY breaking scale $\sqrt F$ and the stop mass $m_{\st}$. }
\label{fig-sqrtF}
\end{figure}

In this paper we will focus on the situation in which the stop decays promptly. More generally, the lifetime of the stop is dominantly a function of its mass, the SUSY-breaking scale $\sqrt{F}$ (equivalently, the gravitino mass), and various Standard Model parameters. An approximate analytical expression for the stop decay rate for $m_\st < m_t$ is~\cite{Chou:1999zb}
\be
\Gamma \sim \frac{\alpha}{\sin^2\theta_W}\frac{\l(m_\st - m_W\r)^7}{128\pi^2 m_W^2 F^2}
\ee
while for $m_\st > m_t$ the decay process gradually starts being dominated by $\st\to t\,\gr$ (with a subsequent $t\to W^+b$ decay) which has the rate~\cite{Sarid:1999zx}
\be
\Gamma = \frac{m_\st^5}{16\pi F^2}\l(1 - \frac{m_t^2}{m_\st^2}\r)^4
\ee
Contours of constant stop lifetime are shown in figure~\ref{fig-sqrtF}.
We see that, as is generally the case in gauge mediation, the lifetime of the NLSP can range from prompt (corresponding to lower SUSY-breaking scales and/or heavier stops) to detector-stable (higher SUSY-breaking scales and/or lighter stops). For $m_\st \lesssim m_t$, prompt decay of the stop requires the SUSY breaking scale $\sqrt F$ to be as small as it can possibly be, on the order of $10$~TeV.

It is also important to consider longer lived stops, but we will not do so in detail in this paper. Stops that are sufficiently stable that they travel fully through the detector are constrained by searches for stable charged or colored particles. The current best limit on detector-stable stops comes from ATLAS and corresponds to $m_\st \gtrsim 300$~GeV~\cite{Aad:2011yf}. (Very long-lived stop NLSPs may also have important consequences for BBN~\cite{DiazCruz:2007fc,Kohri:2008cf}.) Even more interesting is the intermediate case of a stop that decays at a displaced vertex. As far as we know, there are currently no limits on this scenario. This would give rise to signatures involving displaced jets and leptons. This could pose interesting challenges for triggering and reconstruction, as was recently discussed in a related context in~\cite{Meade:2010ji}.

\begin{figure}[t]
\begin{center}
\includegraphics[width=0.4212\textwidth]{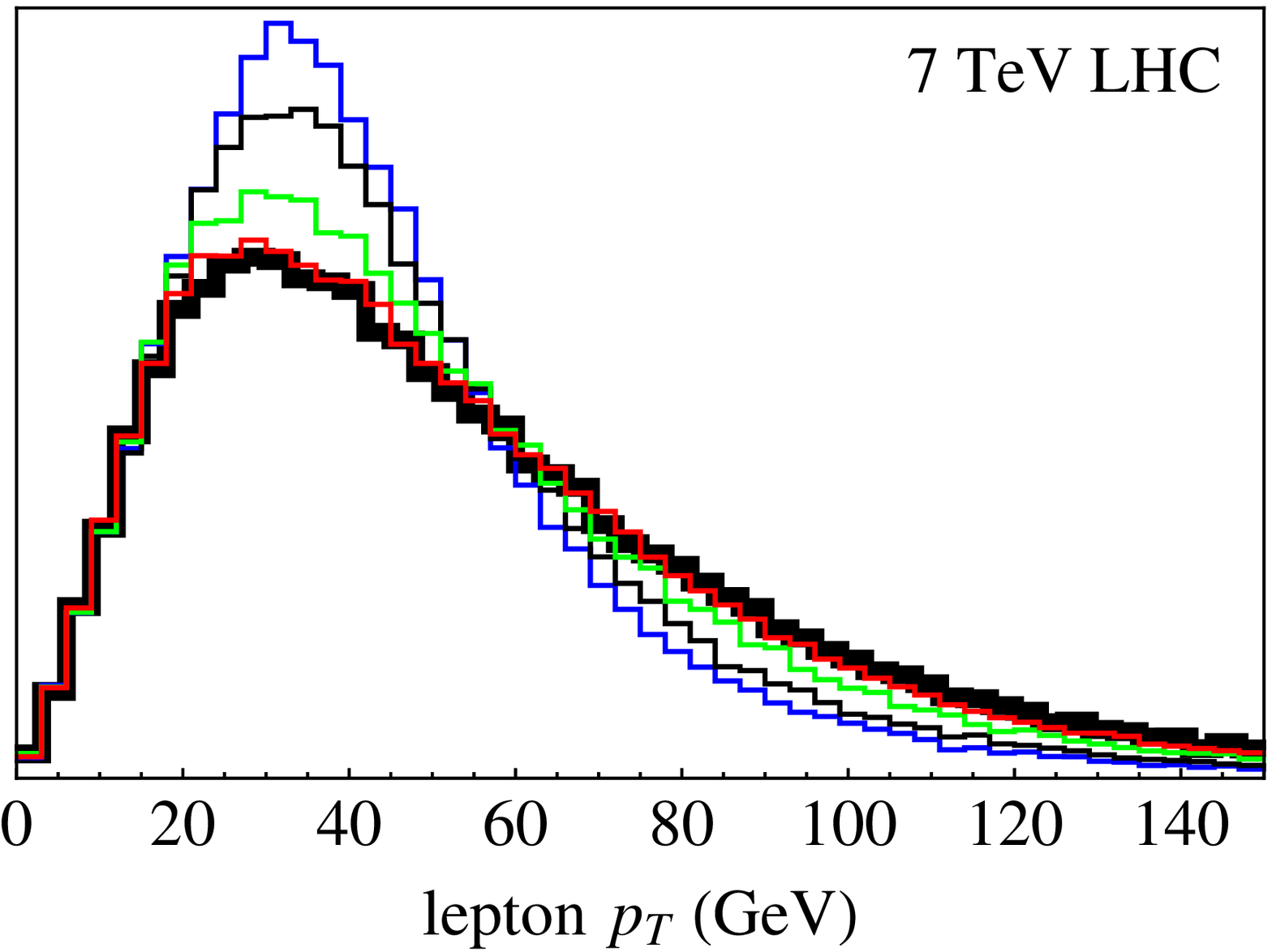}\q
\includegraphics[width=0.4312\textwidth]{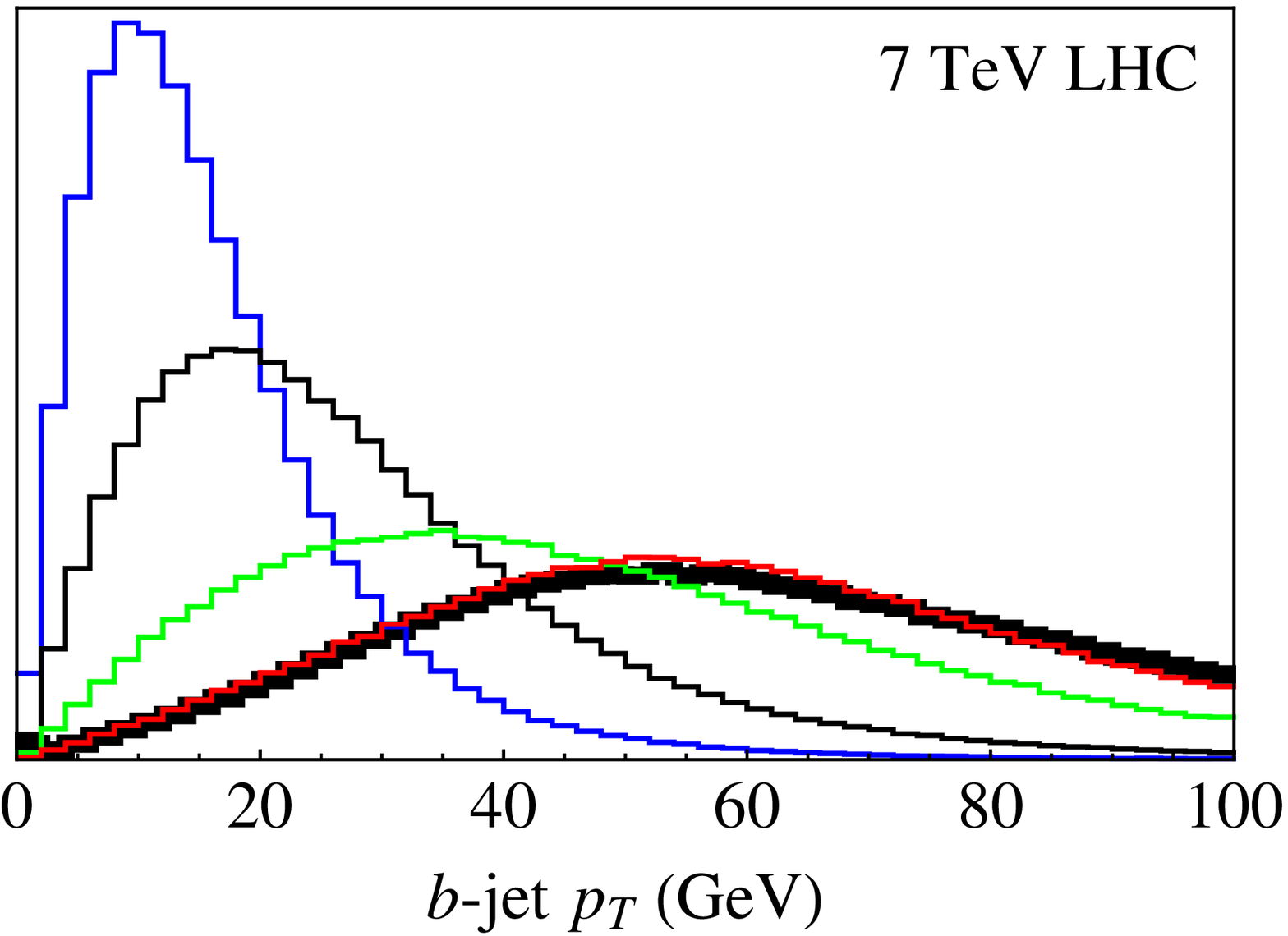}\\\vspace{5mm}
\includegraphics[width=0.3012\textwidth]{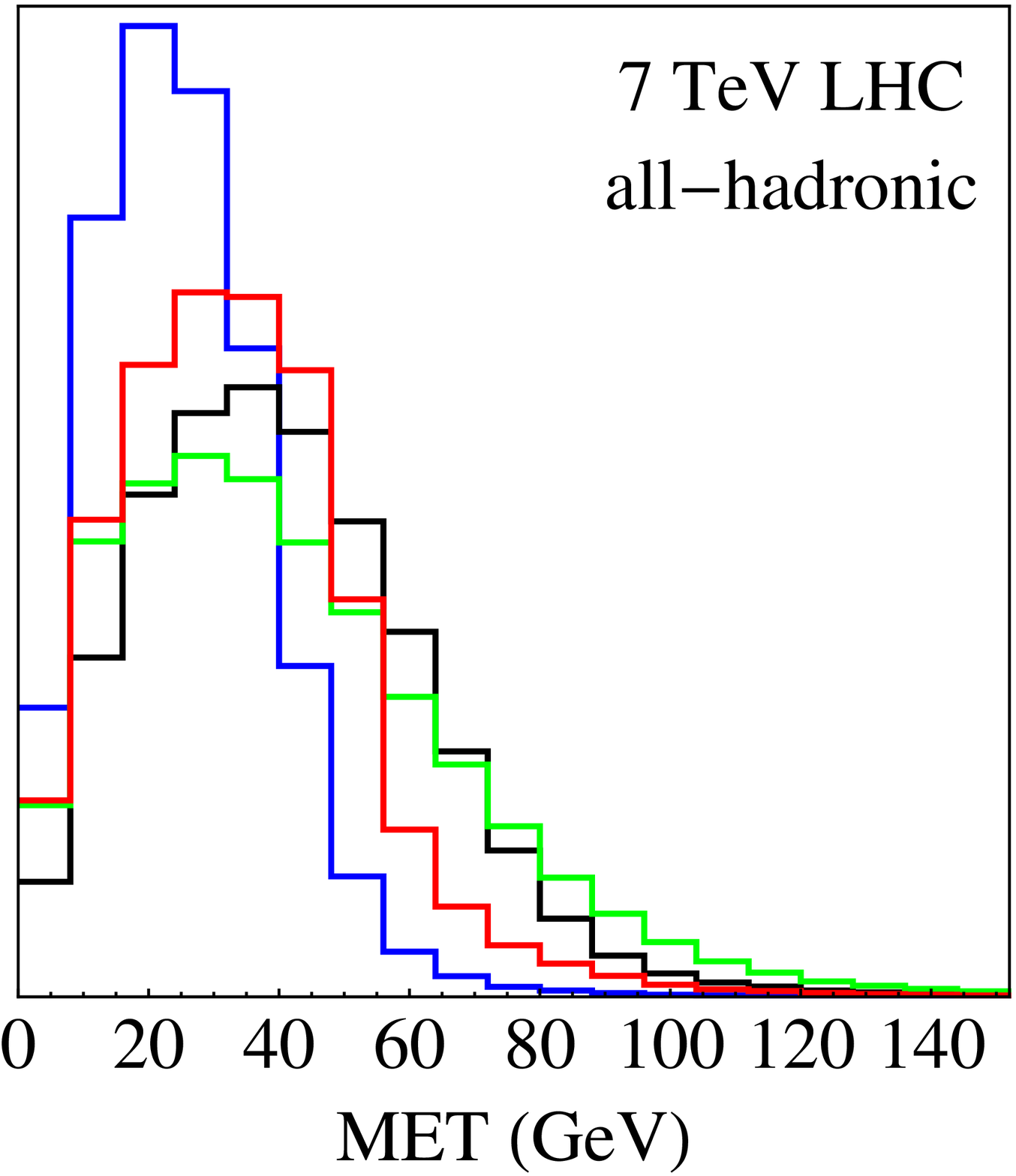}\q
\includegraphics[width=0.3012\textwidth]{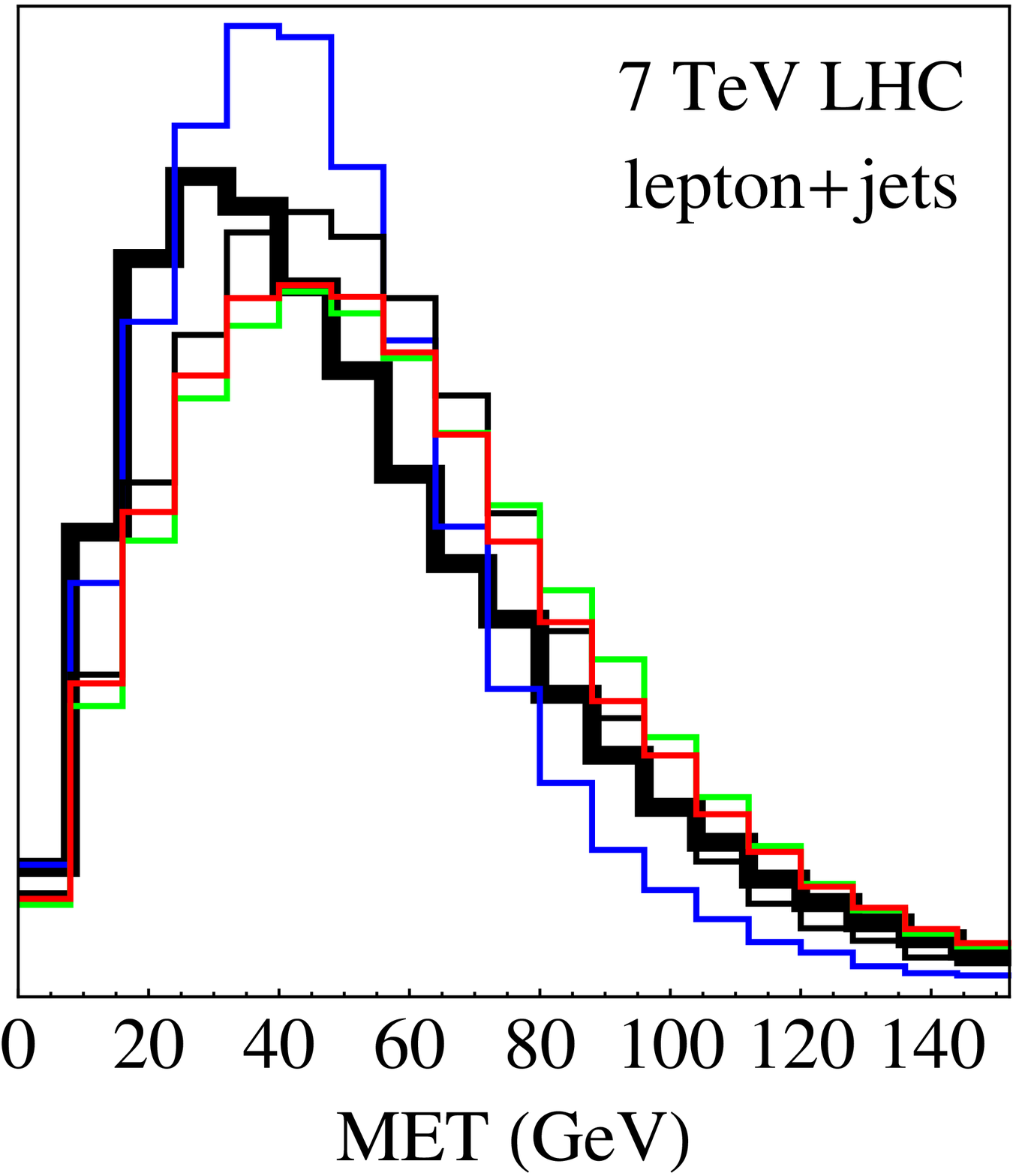}\q
\includegraphics[width=0.3012\textwidth]{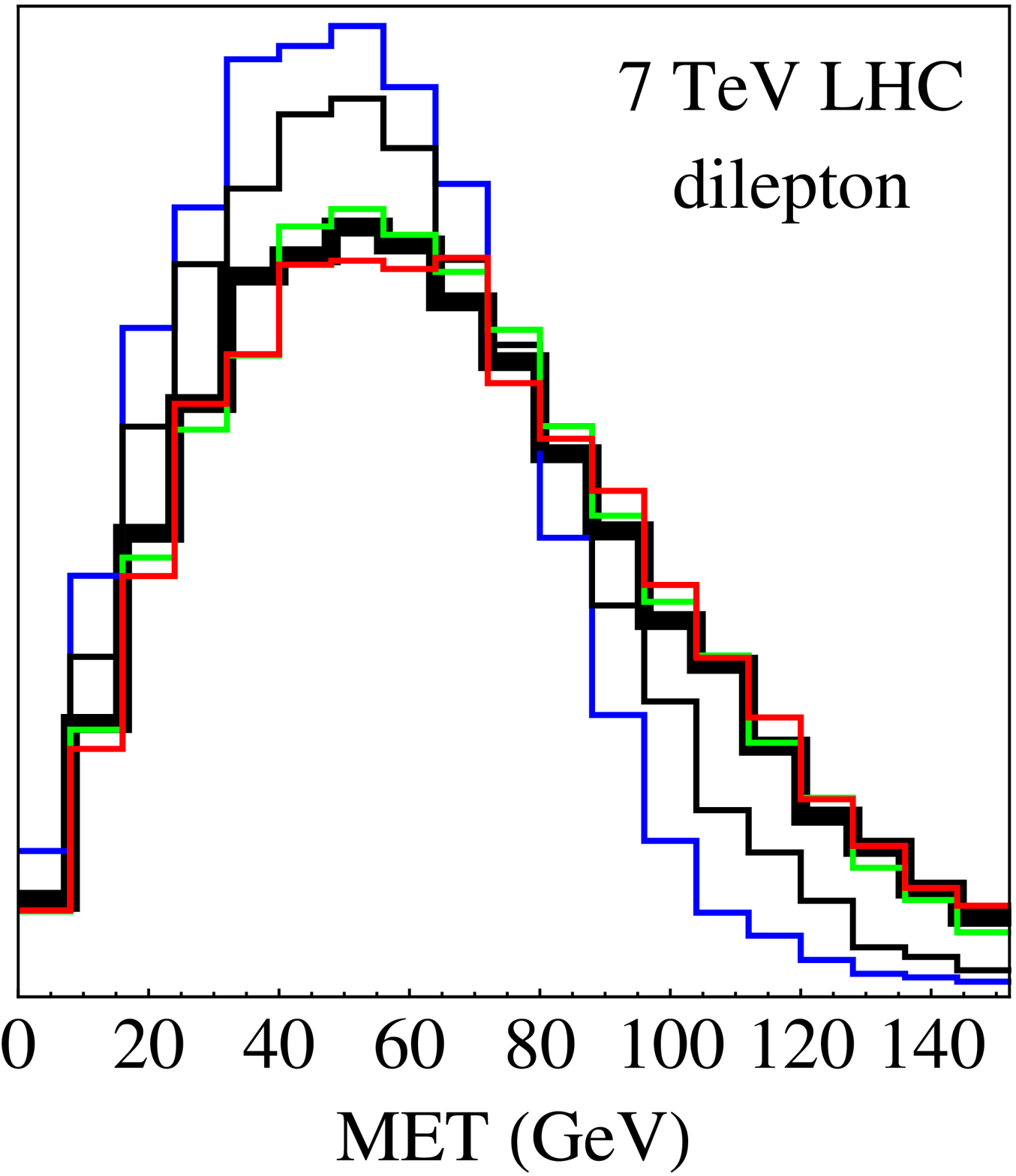}
\end{center}
\caption{Parton-level distributions of the leptons $p_T$ (top left), $b$-jets $p_T$ (top right) and $\slash E_T$ in the different channels (bottom) for $t\bar t$ (thick black line) and stops with masses (in GeV) $120$ (blue), $150$ (black), $180$ (green) and $200$ (red).}
\label{fig-pT-LHC}
\begin{center}
\includegraphics[width=0.4212\textwidth]{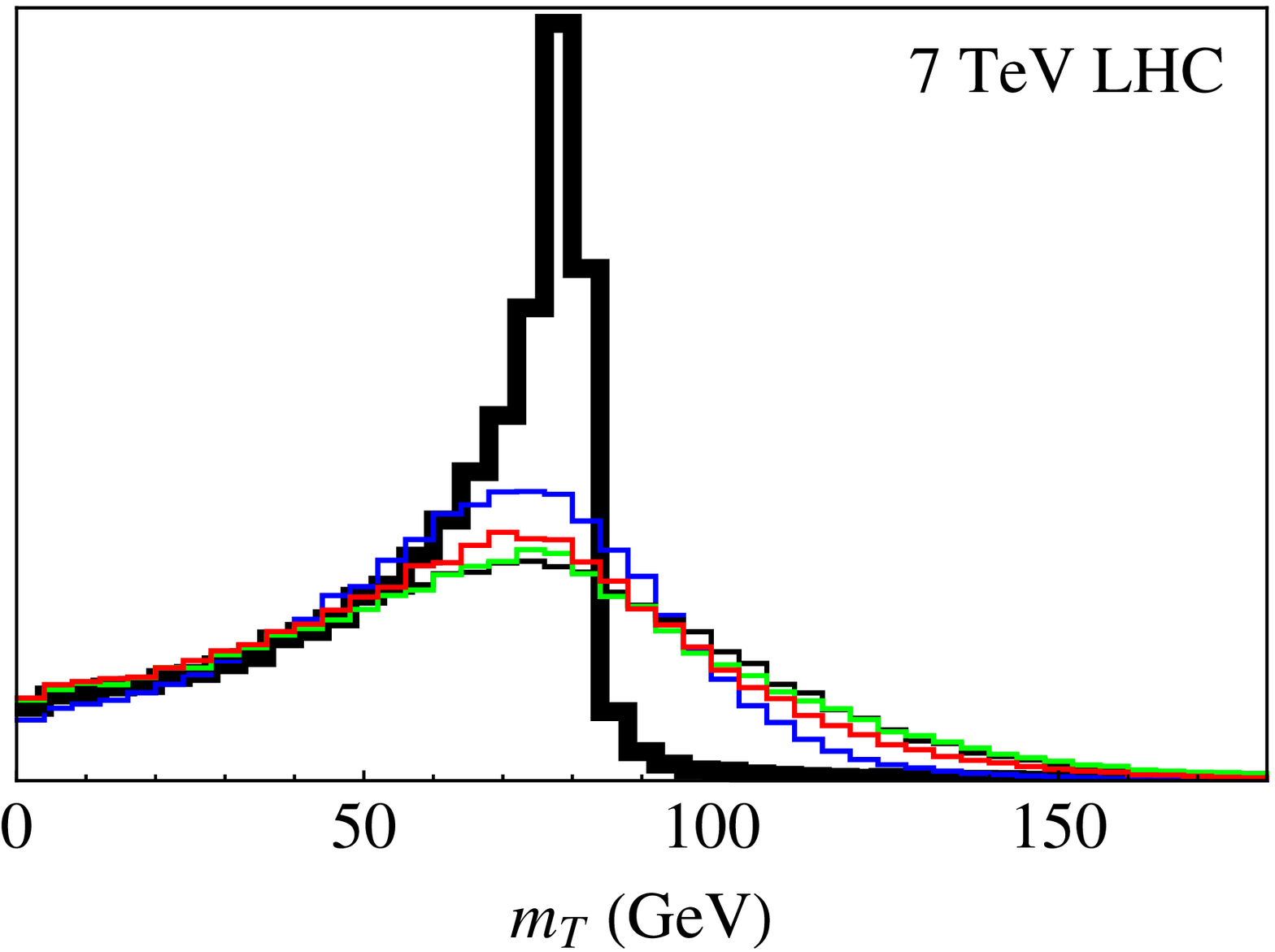}
\includegraphics[width=0.4212\textwidth]{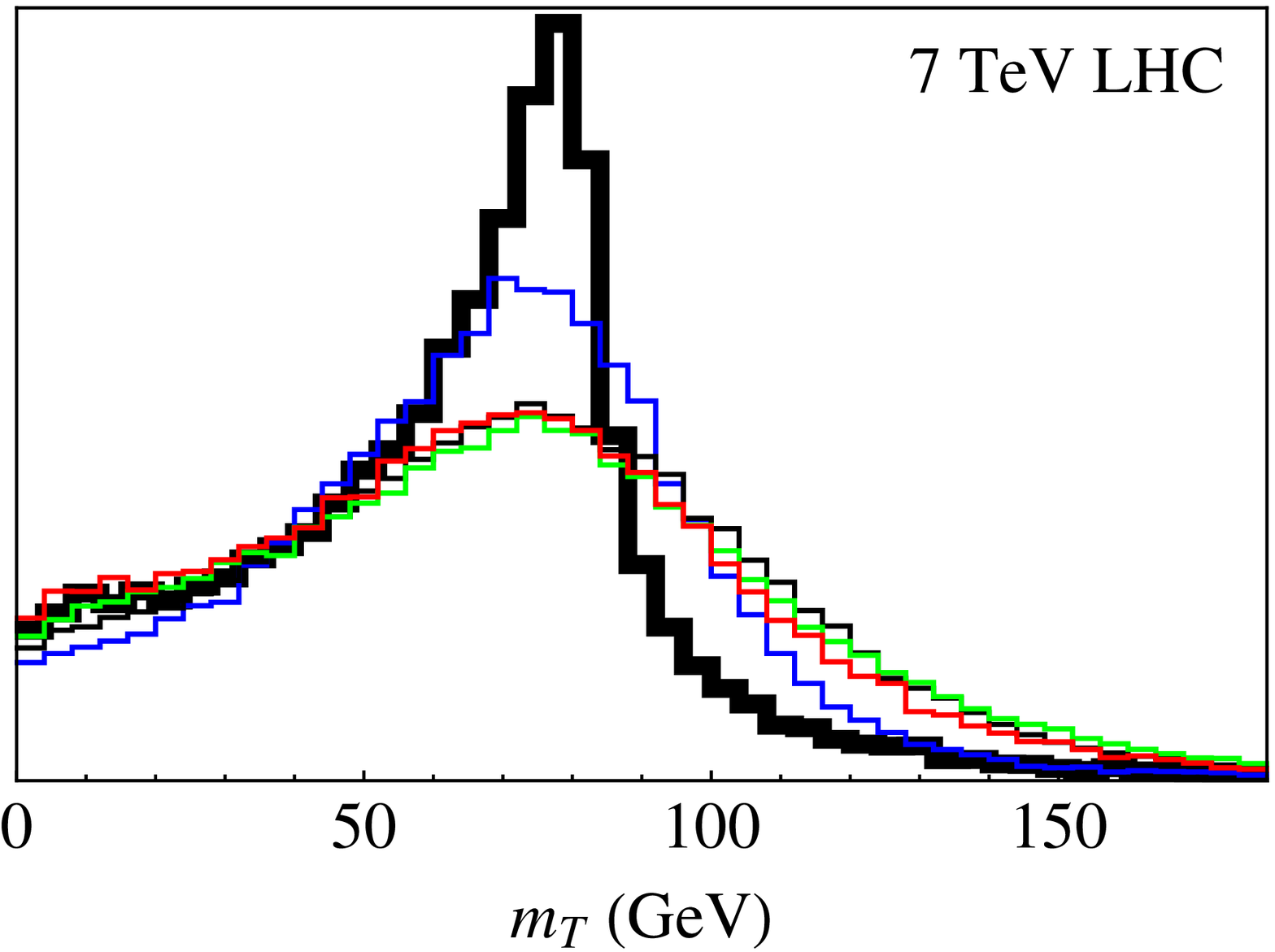}
\end{center}
\caption{Left: parton-level distributions of $m_T$, eq.~(\ref{mT-defn}), in the lepton+jets channel. Right: same distributions after showering (the full $t\bar t$ sample) and applying geometric acceptance, lepton selection and the dilepton veto (as defined in~\cite{ATLAS-CONF-2011-036}).}
\label{fig-mT-LHC}
\end{figure}

\begin{figure}[t]
\begin{center}
\includegraphics[width=0.4212\textwidth]{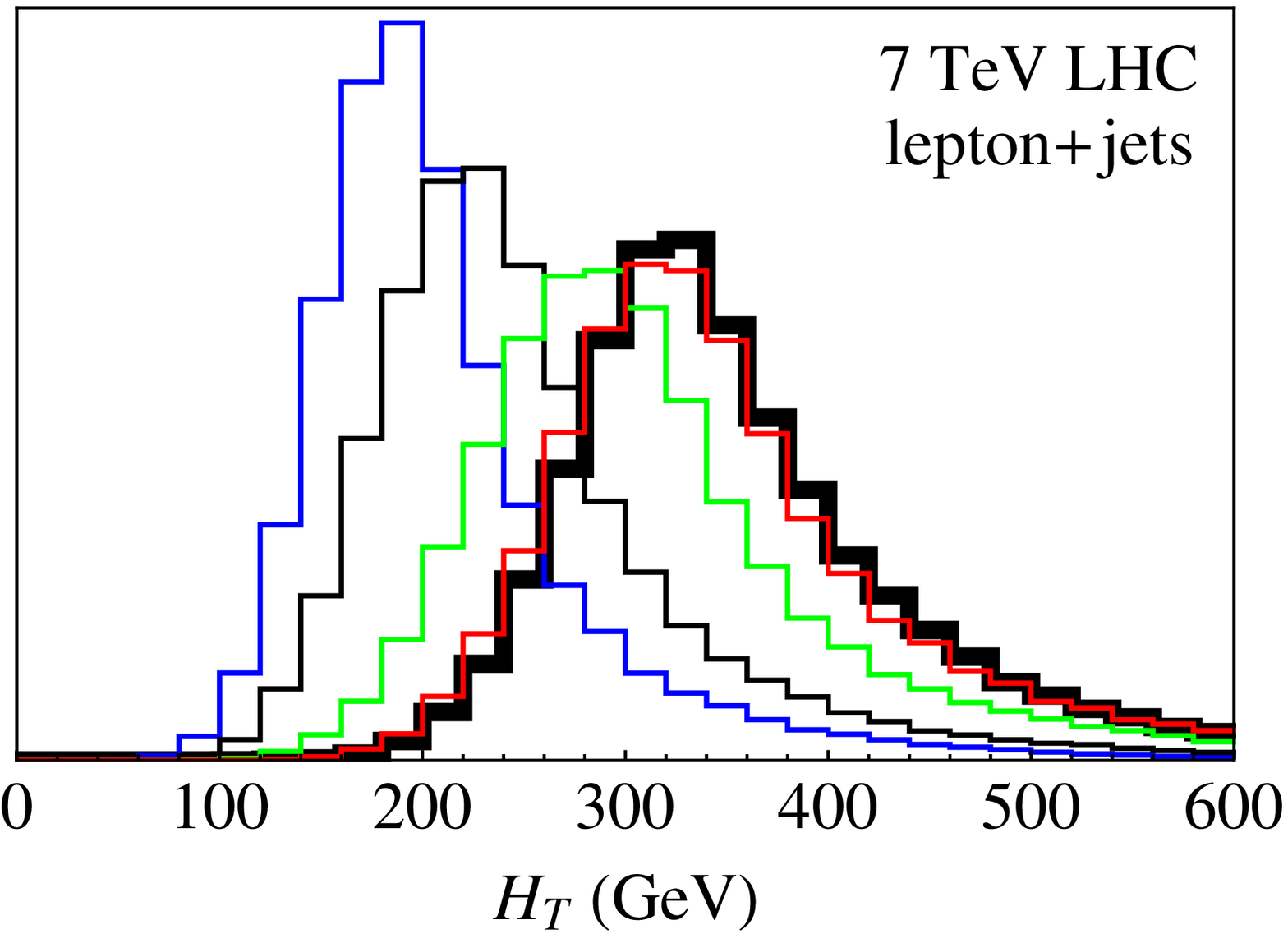}\q
\includegraphics[width=0.4212\textwidth]{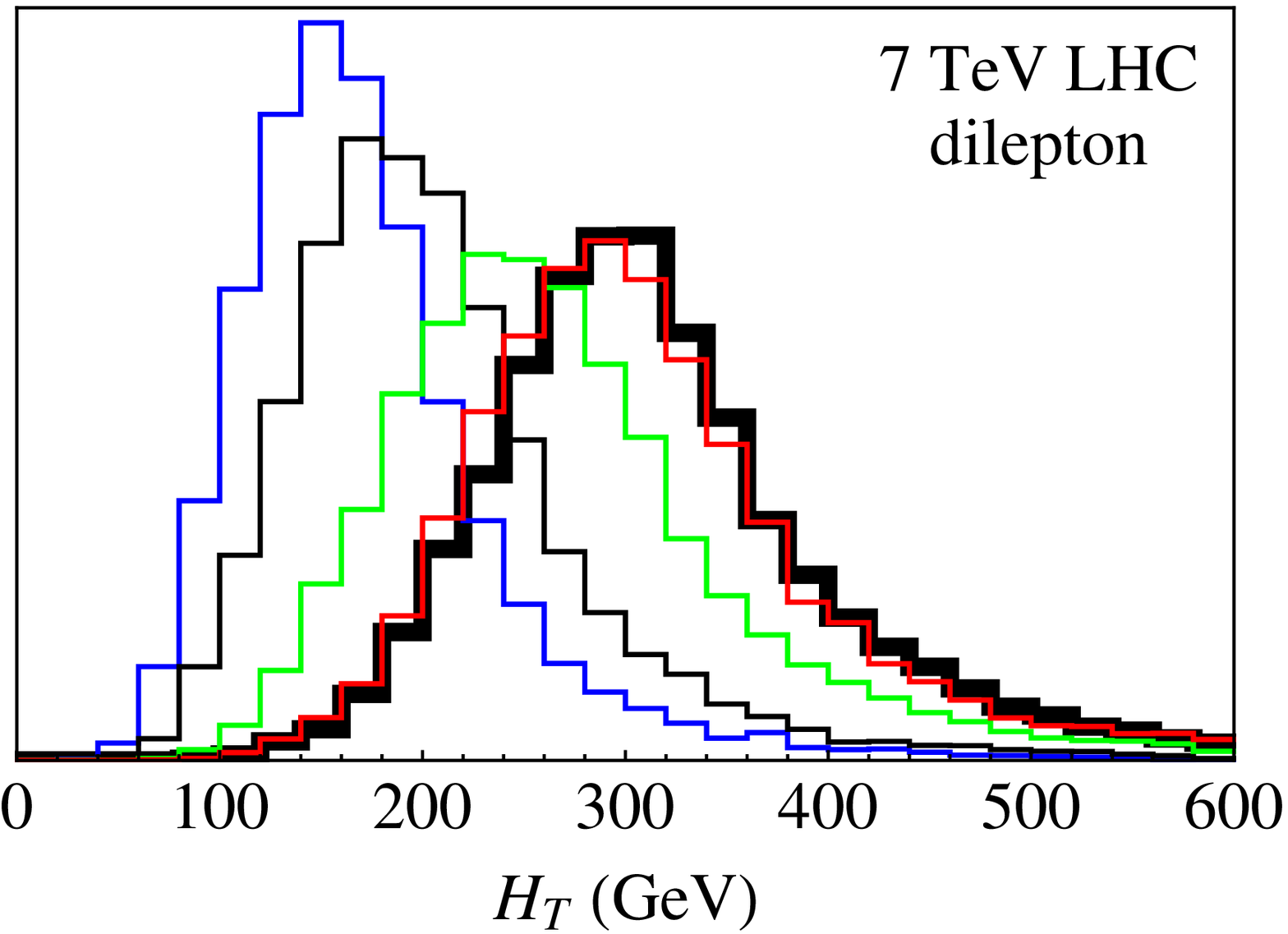}
\end{center}
\caption{Parton-level distributions of $H_T$, eq.~(\ref{HT-defn}), for the lepton+jets (left) and dilepton channel (right).}
\label{fig-HT-LHC}
\end{figure}

\begin{figure}
\begin{center}
\includegraphics[width=0.3012\textwidth]{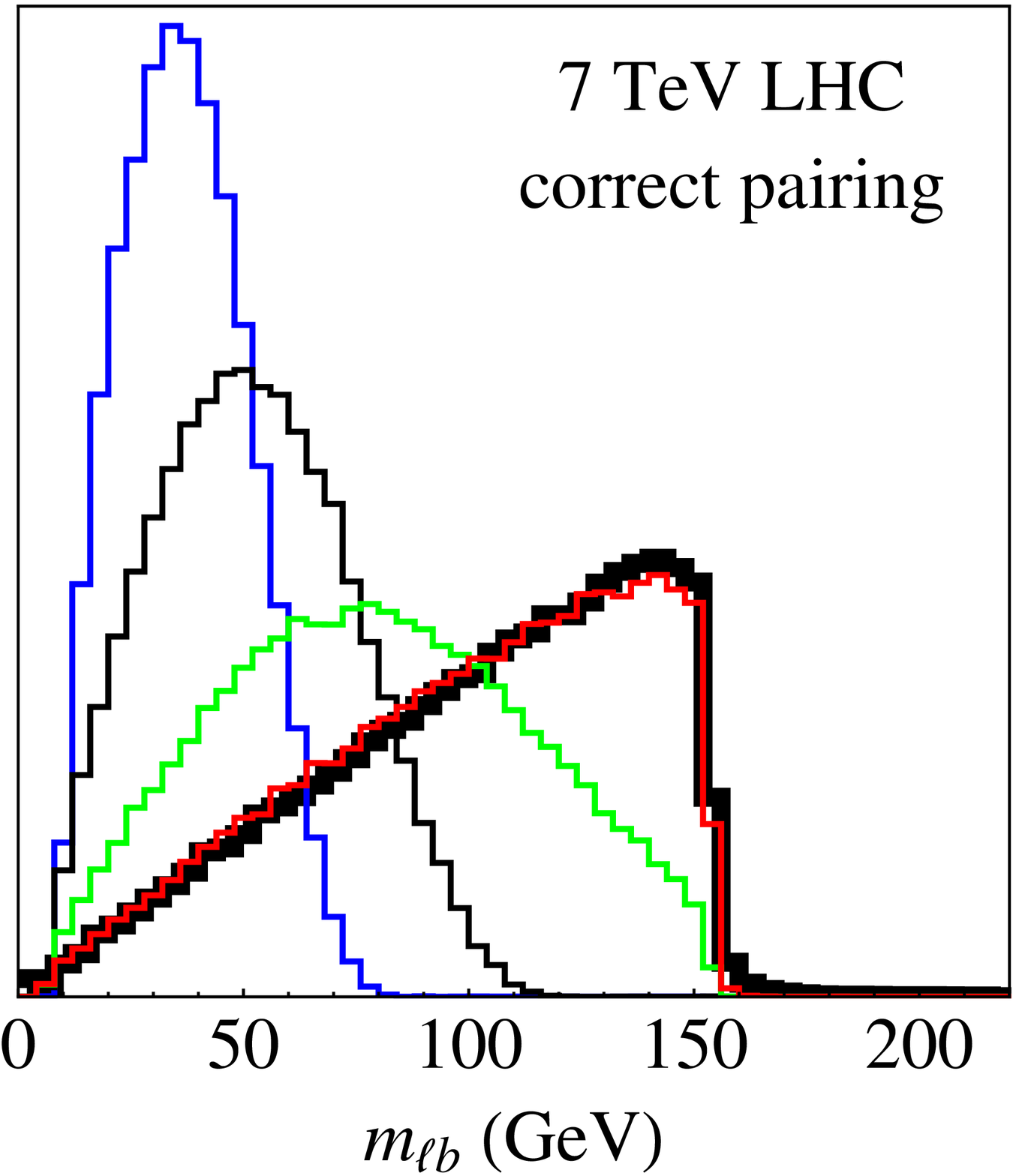}\q
\includegraphics[width=0.3012\textwidth]{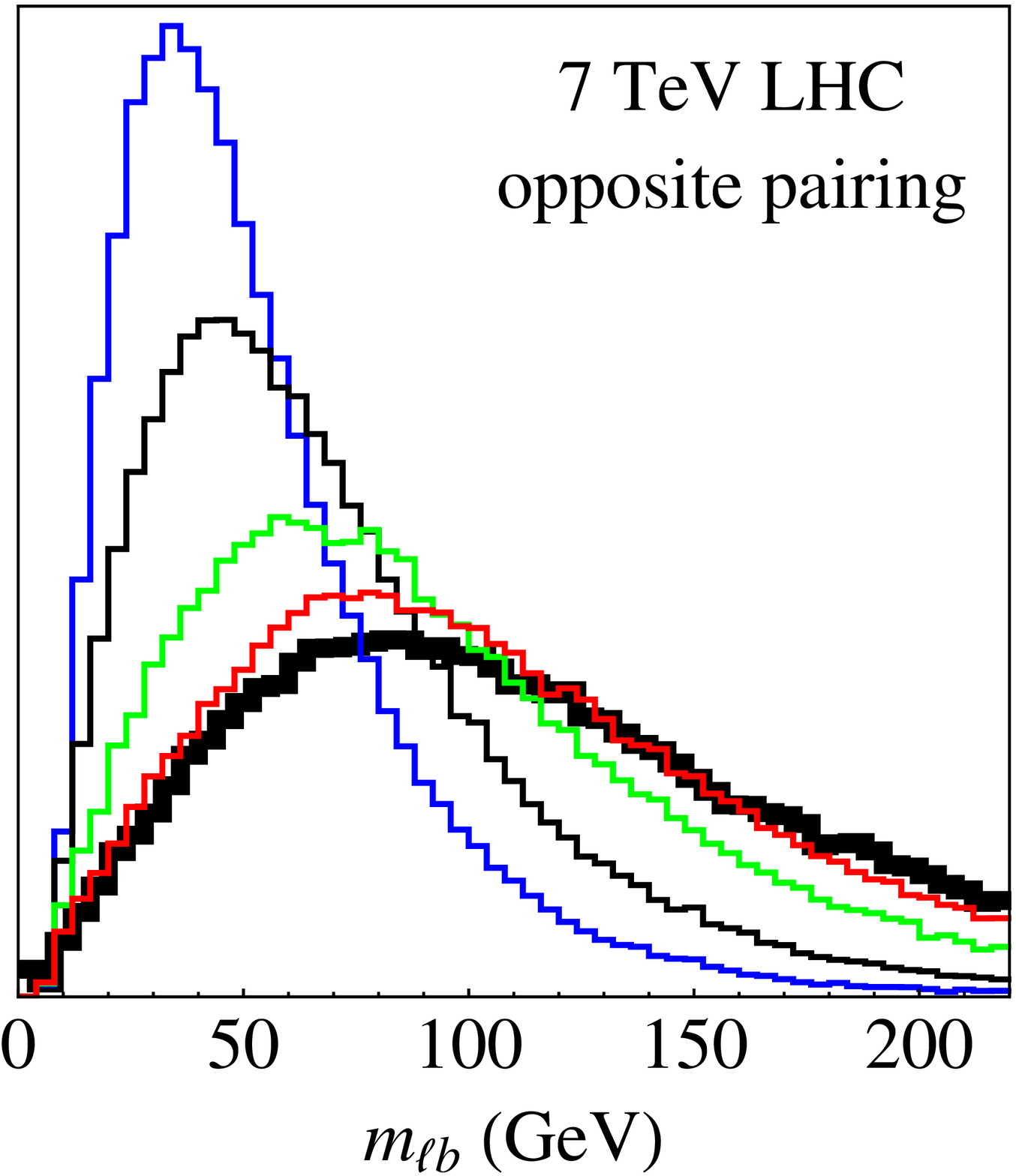}\q
\includegraphics[width=0.3012\textwidth]{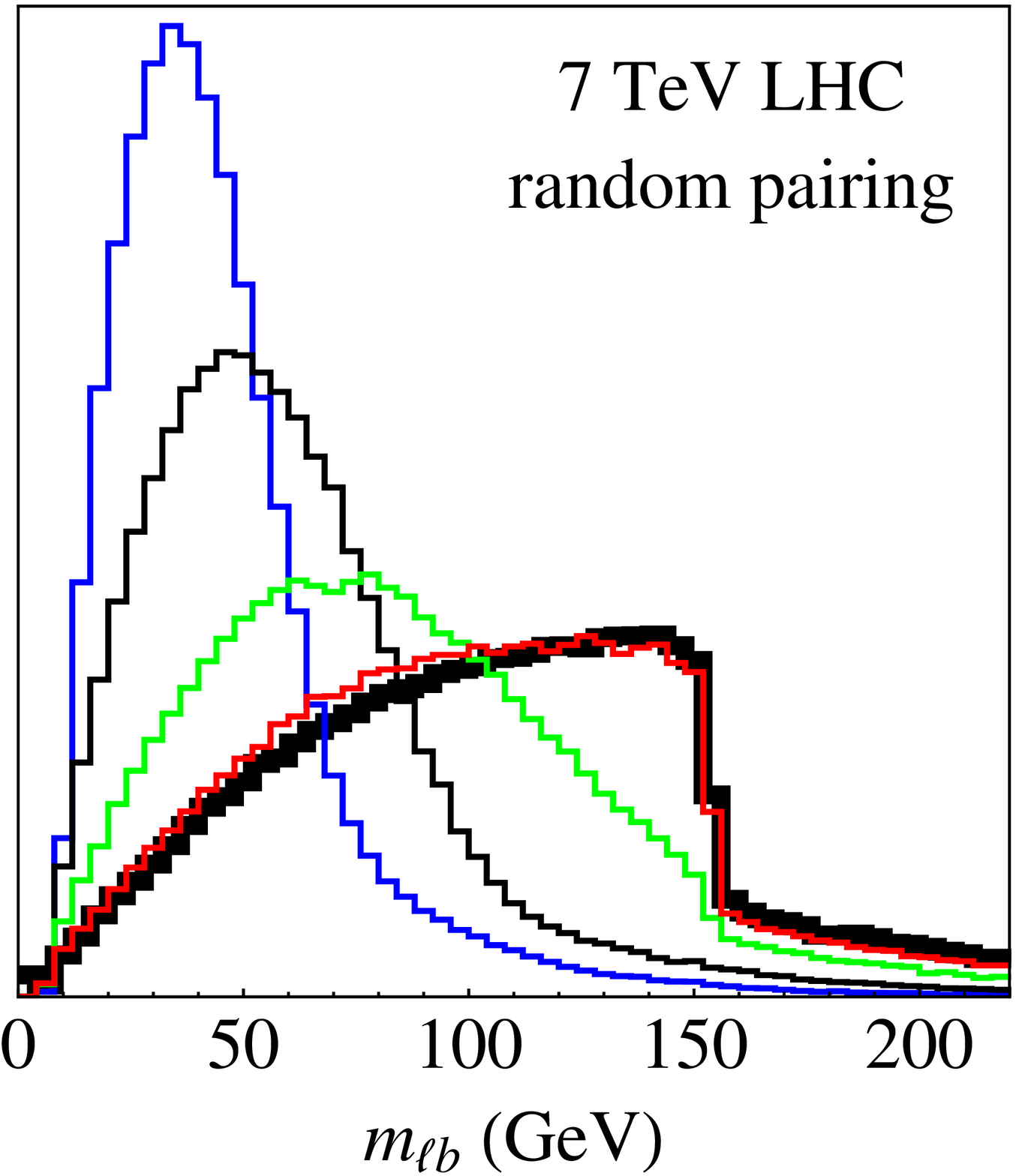}
\end{center}
\caption{Parton-level distributions of $m_{\ell b}$ where the lepton and the $b$ quark come from the same top or stop (left), the opposite combination (middle) and random pairing (right).}
\label{fig-mlb-LHC}
\end{figure}

\subsection{Kinematic distributions}
\label{sec-kin-dist}

In figures~\ref{fig-pT-LHC}--\ref{fig-mlb-LHC}, we plot the distributions of  various kinematic quantities characterizing pair-production of stop NLSP. These plots were made using a combination of our own code for the decay of $\st\to Wb\gr$ according to the matrix element~(\ref{ME}); and {\sc Pythia} for everything else.\footnote{Note that decaying the $W$ through {\sc Pythia} has the effect of averaging over the $W$ polarization. This is not a problem since the $W$ polarization fractions for the stop are very model dependent anyway, while their effect on the basic kinematic distributions is relatively small~\cite{Chou:1999zb}.} Unless stated explicitly, these distributions are purely parton-level, i.e.\ they do not include showering, hadronization, detector simulation, or any cuts.

The distributions for the Tevatron and the $7$~TeV LHC turn out to be very similar, both qualitatively and quantitatively (this happens because the tops and stops are produced relatively close to threshold), so we present them only for the LHC case. For comparison, we have also included distributions of the same quantities for $t\bar t$.

The transverse mass $m_T^2 = E_T^2 - p_T^2$ of the $W$, is determined from the measured quantities in the lepton+jets channel as would be appropriate if the neutrino were the only invisible particle, namely
\be
m_T = \sqrt{2\l(p_T^\ell \slash E_T - \vv{p}_T^\ell\cdot\vv{\slash E}_T\r)} = \sqrt{2 p_T^\ell \slash E_T\l(1 - \cos\Delta\phi_{\ell,\slash E_T}\r)}
\label{mT-defn}
\ee
$H_T$ is defined as\footnote{Note though that the definition of $H_T$ varies between the different experimental analyses, sometimes excluding $\slash E_T$ or including only jets. The definition (\ref{HT-defn}) which we use in figure~\ref{fig-HT-LHC} is often also denoted $m_{\rm eff}$.}
\be
H_T = \sum_{\rm leptons} p_T + \sum_{\rm jets} p_T + \slash E_T
\label{HT-defn}
\ee
Some of the distributions are shown separately for the different decay possibilities of the two $W$'s: dilepton, lepton+jets and all-hadronic (by lepton, we mean $e$ or $\mu$). Note that unlike for $t\bar t$, stop events have significant missing transverse energy (MET) carried by the gravitinos even when both $W$'s decay hadronically, which results in a jets+MET signature.

Many of these kinematic quantities are used in the experimental analyses that we will study in this paper, either to select for $t\bar t$ events, or to distinguish such events from events with stops or other top partners. One major exception is $m_{\ell b}$, the invariant mass of the lepton and the $b$-quark coming from top or stop decay. The $m_{\ell b}$ distributions are not used in any of the existing analyses, even though they can be extremely useful in distinguishing between stop and top events~\cite{Chou:1999zb}. We will return to $m_{\ell b}$ in section~\ref{sec-mlb}.

One of the most significant differences relative to the top for $m_{\st}\lesssim m_t$ is that the $b$ jets are much softer (figure~\ref{fig-pT-LHC}). This happens due to purely kinematic reasons: in the decay $t\to W^+b$, the $b$ momentum (in the top rest frame) is
\be
p_b = \frac{m_t}{2}\l(1 - \frac{m_W^2}{m_t^2}\r)
\ee
(where for simplicity we neglected $m_b$), which creates a Jacobian peak in $p_{b,T}$ (which gets somewhat smeared, as seen in figure~\ref{fig-pT-LHC}, because the tops are produced with finite $p_T$). On the other hand, in the case of the 3-body decay $\st\to W^+b\gr$ the momenta of the $b$ quarks are distributed within
\be
p_b \leq \frac{m_\st}{2}\l(1 - \frac{m_W^2}{m_\st^2}\r)
\ee
Additional differences are a significant high-$m_T$ tail (in the lepton+jets channel) and the somewhat lower values of $H_T$ in the lepton+jets and dilepton channels. In section~\ref{sec-results} we will see how these differences determine the acceptance of the stops in the various existing analyses.

\section{Overview of relevant Tevatron and LHC analyses\label{sec-stop-searches}}
\setcounter{equation}{0}

In this section, we will give a brief but comprehensive overview of the Tevatron and LHC analyses which are relevant for constraining the stop NLSP scenario. We have divided up the discussion into two categories: Standard Model analyses of $t\bar t$ production, where stop pair production could be ``hiding''; and explicit searches for new physics with stop-like final states, for which $t\bar t$ is a major background.

\subsection{$t\bar t$ analyses}

\begin{table}[t]
\begin{center}
\begin{tabular}{|c|c|c|c|c|c|}\hline
Analysis & Leptons & $b$ tags & Luminosity & $\begin{array}{c} \mbox{Excluded} \\ \mbox{stop masses} \end{array}$ & \\\hline\hline
CDF~\cite{CDF-9890}             & 2 & $\geq 0$ & 4.47 fb$^{-1}$ & - & \checkmark \\
                                &   & $\geq 1$ & 4.47 fb$^{-1}$ & - & \\\hline
CDF~\cite{CDF-10163}            & 2 & $\geq 0$ & 5.1 fb$^{-1}$  & - & \\
                                &   & $\geq 1$ & 4.8 fb$^{-1}$  & - & \\\hline
D0~\cite{Abazov:2011cq}         & 2 & $\geq 0$ & 5.4 fb$^{-1}$  & \textit{not simulated} & \\\hline
CDF~\cite{Aaltonen:2010ic}      & 1 & $\geq 1$ & 4.3 fb$^{-1}$  & - & \\\hline
D0~\cite{Abazov:2011mi}         & 1 & $\geq 1$ & 5.3 fb$^{-1}$  & \textit{not simulated} & \\\hline
\hline
ATLAS~\cite{ATLAS-CONF-2011-034}& 2 & $\geq 0$ & 35 pb$^{-1}$   & - & \checkmark \\
                                &   & $\geq 1$ & 35 pb$^{-1}$   & - & \\\hline
CMS~\cite{CMS-TOP-11-002}       & 2 & $\geq 0$ & 36 pb$^{-1}$   & \textit{not simulated} & \\
                                &   & $\geq 1$ & 36 pb$^{-1}$   & \textit{not simulated} & \\\hline
ATLAS~\cite{ATLAS-CONF-2011-023,ATLAS-CONF-2011-035}
                                & 1 & $\geq 0$ & 35 pb$^{-1}$   & - & \\
                                &   & $\geq 1$ & 35 pb$^{-1}$   & - & \\\hline
\end{tabular}
\end{center}
\caption{The most recent Tevatron and LHC $t\bar t$ cross section measurements (with cut-and-count analyses). The dashes indicate that no ranges of stop NLSP masses were found to be excluded at $95\%$ CL by our reproduction of the analyses. The last column indicates the most sensitive analyses whose results will be presented in more detail.}
\label{tab-tt-analyses}
\end{table}

Because of the general similarity between top and stop NLSP signatures,  measurements of the $t\bar t$ cross section have the potential to also constrain stop pair production. Experimental analyses of $t\bar t$ production fall into four categories, based on how the two $W$'s decay: dileptonic (5\%), lepton+jets (35\%), jets+MET (10\%), or all hadronic (50\%).\footnote{By lepton we mean $e$ or $\mu$ (including the case of a leptonically decaying $\tau$). The jets+MET category consists of zero-lepton events where at least one $W$ decays to $\tau\nu$ and the $\tau$ subsequently decays hadronically. The all-hadronic category, by contrast, consists of events with no intrinsic MET, where both $W$'s have decayed hadronically.} The dilepton channel is generally the cleanest: after requiring two leptons, two jets, and MET, the signal-to-background ratio for $t\bar t$ events is typically $\sim 2$, and after requiring at least one of the jets to be $b$-tagged it increases to $\sim 10$. However, due to the small branching ratio of this mode there is statistical uncertainty roughly of the same size as the systematic uncertainty (this is true for both $5$~fb$^{-1}$ at the Tevatron and $35$~pb$^{-1}$ at the LHC, for both pre-tag and $b$-tagged samples). The lepton+jets channel has a much larger branching ratio, however the pre-tag sample is dominated by the $W$+jets background, so usually only the $b$-tagged sample is relevant (unless more sophisticated techniques that assume detailed information about the signal properties are used, but those would have less sensitivity to new physics). After a typical set of requirements (a lepton, $\geq 4$ jets, MET) the signal-to-background ratio is $\sim 4$ and the systematic uncertainty dominates over the statistical uncertainty (by a factor of $\sim 2$).
The most recent Tevatron and LHC $t\bar t$ cross section measurements in the dilepton and lepton+jets final states using cut-and-count methods are listed in table~\ref{tab-tt-analyses}.

To get a rough idea of the sensitivity of $t\bar t$ cross section measurements to stop pair production in these channels, we can consider the ratio of stop and top production cross sections, which can be read off figure~\ref{fig-stop-cs}. For example, for $m_\st = m_t$, $\sigma_{\st\st^\ast}/\sigma_{t\bar t} = 0.10$ at the Tevatron and $0.18$ at the LHC. Since $t\bar t$ is a dominant background to stop pair production, this gives a crude estimate for the signal-to-background ratios, both with and without $b$ tagging. Of course, these numbers will be affected, sometimes very significantly, by the differences in the kinematic properties between stop and top events discussed in section~\ref{sec-stop-NLSP}.

Decay modes without leptons seem less useful because of the large QCD background. A possible exception is the  jets+MET channel. Because of the MET carried by the gravitinos, most of the stop events fall in this category. By contrast, the $t\bar t$ background is significantly reduced (branching ratio $\sim 10\%$ as indicated above). The QCD background (where MET can arise due to mismeasured jets) is still an obstacle for a cut-and-count analysis, but a dedicated search using for example neural networks may work (similarly to the top mass measurement in this channel~\cite{CDF-10433}). However, designing such a search and examining its feasibility is beyond the scope of the present paper.

Various properties of the top quark have been measured besides the cross section. In particular, the top mass has been measured at the Tevatron with the great precision of $1$~GeV. However, these measurements are not necessarily more sensitive to the presence of stop events. Many of them use matrix elements or neural networks for background rejection and are therefore likely to discriminate against stop events. Others, in order to extract the desired quantity and/or obtain a pure sample of tops, require a detailed reconstruction of the $t\bar t$ event from the observed objects. Since the momentum carried by the gravitinos will be unaccounted for, stop events will not reconstruct in a meaningful way and are likely to be rejected as background or make a relatively smooth contribution to the measured distributions. These analyses also present a practical difficulty for us, since without having access to the code used in the experimental study, reproducing the behavior of  complex algorithms for events which they were not designed to treat would be risky. For these reasons, we do not include these measurements in our study.

\subsection{Searches for new physics}

Several experimental studies have looked specifically for light stops, but not in the GMSB scenario with stop NLSP that we consider here. However, in some cases the stops also had $t\bar t$-like signatures and such analyses can have good acceptance to GMSB stops as well (despite the fact that they were optimized for other scenarios). We have listed the relevant searches in table~\ref{tab-np-analyses}, and will now proceed to describe them in more detail.

At the Tevatron, CDF~\cite{Aaltonen:2010uf} and D0~\cite{Abazov:2008kz,D0-5937,Abazov:2010xm} have considered the situation in which the stop decays via a virtual chargino as
\be
\st\to b\ell^+\tilde\nu
\label{sneutrino}
\ee
where the massive sneutrino $\tilde\nu$ can be the LSP or decay invisibly as $\tilde\nu \to \nu\tilde\chi_1^0$ or $\nu\gr$. The signature here is similar to that of the dilepton channel of (\ref{stop-decay}). The D0 search~\cite{Abazov:2010xm} (with $5.4$~fb$^{-1}$ of data) used distributions of composite discriminant variables optimized for the decay (\ref{sneutrino}) as a function of the stop-sneutrino mass difference, and excluded stops with masses as large as $240$~GeV for a certain range of $\tilde\nu$ masses (assuming $100\%$ dilepton branching ratio, an order of magnitude larger than in our case).

Another possibility that has been studied is
\be
\st\to b\tilde\chi_1^+ \,,\qq
\tilde\chi_1^+ \to W^{+(\ast)}\tilde\chi_1^0
\label{toplike-stop}
\ee
which also gives rise to a $t\bar t$-like signature. The most recent such study from D0 in the lepton+jets channel (although with only $0.9$~fb$^{-1}$) did not yield significant exclusion limits~\cite{Abazov:2009ps}.

Finally, CDF~\cite{Aaltonen:2009sf,CDF-9439,Johnson:2010zza} has considered a scenario which has some overlap with (\ref{toplike-stop}) in the dileptonic final state:
\be
\st\to b\tilde\chi_1^+ \,,\qq
\tilde\chi_1^+ \to \ell^+\nu\tilde\chi_1^0
\label{stop-chargino-neutralino}
\ee
The CDF study treated the chargino branching ratio to leptons as a free parameter. If the chargino decays through a $W$ as in~(\ref{toplike-stop}), then the dilepton channel will have branching ratio of $0.11$ (as is the case for our stop NLSP scenario). The chargino can also decay through a charged higgs, a slepton or a sneutrino, in which case the branching ratio can be as high as $1$. In order to discriminate between stop and top events, CDF has designed an algorithm for approximately reconstructing the stop mass. Because the final state contains four undetectable particles and the masses of the intermediate chargino and the final neutralino are unknown, it is impossible to reconstruct the event rigorously. Nevertheless, after assuming a mass for the chargino, the algorithm is able to construct a quantity, ``stop mass,'' whose distribution is peaked roughly at the stop mass. For top events, one obtains a much broader distribution peaked somewhat above the top mass. Using $2.7$~fb$^{-1}$ of data, CDF were able to exclude stop masses of up to $200$~GeV for certain chargino and neutralino masses assuming a dilepton branching ratio of $1$, but only a small range of stop masses around $130$~GeV for a branching ratio of $0.11$.

\begin{table}[t]
\begin{center}
\begin{tabular}{|c|c|c|c|c|c|}\hline
Analysis & Leptons & $b$ tags & Luminosity & $\begin{array}{c} \mbox{Excluded} \\ \mbox{stop masses} \end{array}$ & \\\hline\hline
D0 stop search~\cite{Abazov:2010xm}                 & 2 & $\geq 0$ & 5.4 fb$^{-1}$  & $m_\st \lesssim 123$ & \checkmark \\\hline
CDF stop search~\cite{Aaltonen:2009sf}              & 2 &      $0$ & 2.7 fb$^{-1}$  & - & \\
                                                    &   & $\geq 1$ & 2.7 fb$^{-1}$  & $105 \lesssim m_\st \lesssim 120$ & \checkmark \\
                                                    & \multicolumn{3}{c|}{with stop mass reconstruction} & $m_\st \lesssim 150$ & \checkmark \\\hline
D0 stop search~\cite{Abazov:2009ps}                 & 1 & $\geq 1$ & 0.9 fb$^{-1}$  & \textit{not simulated} & \\\hline\hline
ATLAS top partner search~\cite{ATLAS-CONF-2011-036} & 1 & $\geq 0$ & 35.3 pb$^{-1}$ & - & \checkmark \\\hline
ATLAS SUSY search~\cite{Aad:2011ks}                 & $\geq 1$ & $\geq 1$  & 35 pb$^{-1}$ & - & \\\hline
\end{tabular}
\end{center}
\caption{Recent new physics searches and the stop NLSP masses (in GeV) excluded by them at $95\%$ CL according to our analysis. The last column indicates the most constraining analyses whose results will be presented in more detail.}
\label{tab-np-analyses}
\end{table}

Among the searches for new physics released by the LHC experiments to date, two may be potentially relevant to light stops. One is the ATLAS SUSY search for events with leptons, $b$-jets and missing energy~\cite{Aad:2011ks}. This search assumed the decay channel (\ref{toplike-stop}). It did not have sufficient exclusion power for directly produced light stop pairs (but only for stops produced in decays of relatively light gluinos). The search was not optimal for such stops, in part because it required the effective mass of the event to be $m_{\rm eff} > 500$~GeV. Another interesting study is an ATLAS search for a heavy top partner (a fermion decaying into a top and an invisible particle)~\cite{ATLAS-CONF-2011-036}. This is potentially better suited to the light stop NLSP scenario, because it does not impose a hard $m_{\rm eff}$ cut.

\section{Constraints from the Tevatron and the LHC\label{sec-results}}
\setcounter{equation}{0}

Using crude detector simulations and object definitions (see appendix~\ref{sec-simulation} for details), we have reproduced a large number of the analyses listed in tables~\ref{tab-tt-analyses} and~\ref{tab-np-analyses} to a reasonable level of accuracy. We have then used this to estimate the current experimental constraints on stop NLSPs. As can be seen from the tables, most analyses currently set no limit at all on stop NLSPs. This is not surprising, given that no analysis has been optimized to this scenario. Experimental sensitivity is degraded by a combination of low acceptance, especially for the lighter stops, and/or low cross section, especially for the heavier stops.

In this section, we will describe in more detail those analyses that we have found to be most sensitive to the stop NLSP scenario. These are indicated by the check marks in tables~\ref{tab-tt-analyses} and~\ref{tab-np-analyses}. As we will discuss below, these searches succeed either because they are more accepting to soft jets, or because they use more discriminating variables such as $m_T$ or the ``reconstructed stop mass''.

For the most part, we have focused on the simple ``cut-and-count'' portions of these analyses. This is the type of analysis that we can simulate most reliably, and anyway, we expect such analyses to be the most receptive  to scenarios that differ somewhat from those for which they were designed. One exception is the CDF stop search~\cite{Aaltonen:2009sf}, where we were able to reproduce their more sophisticated analysis based on the ``stop mass'' reconstruction. This actually ends up setting the best limit on the stop mass. For the D0 stop search~\cite{Abazov:2010xm}, which also uses a more sophisticated approach, we  were unfortunately unable to make use of their distributions of  composite variables without having access to data. Here we will only use the simple cuts which precede that procedure (the ``selection 1'' described in~\cite{Abazov:2010xm}).

We will discuss the ``cut-and-count'' analyses in section~\ref{sec-cut-and-count}, and the CDF stop mass reconstruction in section~\ref{sec-mass-rec}.

\subsection{Cut-and-count analyses}
\label{sec-cut-and-count}

\begin{figure}[t]
\begin{center}
\includegraphics[width=0.47\textwidth]{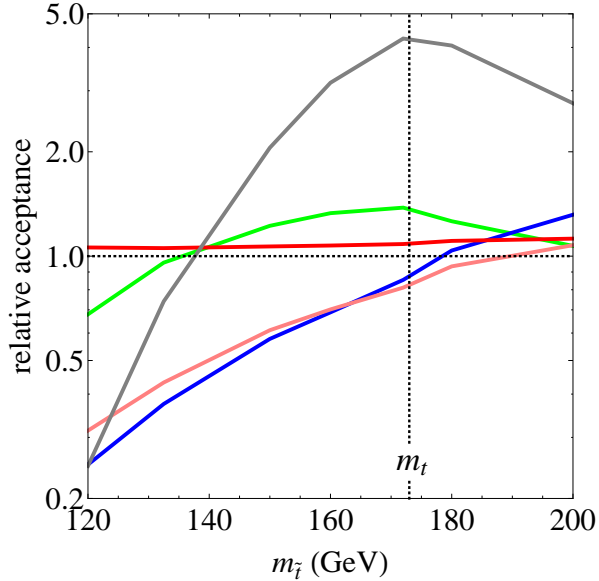}
\end{center}
\caption{Acceptances of $\st\st^\ast$ events relative to those of $t\bar t$ events for the pre-tag sample of the CDF $t\bar t$ cross section measurement in the dilepton channel~\cite{CDF-9890} (blue), the D0 stop search in the $e\mu$ channel (up to selection 1)~\cite{Abazov:2010xm} (red), the $b$-tagged sample of the CDF stop search in the dilepton channel~\cite{Aaltonen:2009sf} (green), the pre-tag sample of the ATLAS $t\bar t$ cross section measurement in the dilepton channel~\cite{ATLAS-CONF-2011-034} (pink) and the ATLAS top partner search~\cite{ATLAS-CONF-2011-036} (gray).}
\label{fig-acc}
\end{figure}

For each analysis and each final state, we compute the expected number of stop events using the following formula:
\be
N_{\st\st^*} = \left({\epsilon_{\st\st^*}\over \epsilon_{t\bar t}}\right) \times \left({\sigma_{\st\st^*}\over \sigma_{t\bar t}}\right) \times N_{t\bar t}
\label{eq-nstop}
\ee
The first factor is obtained from our simulation. It is the acceptance for stop pair production relative to that for $t\bar t$. This is shown in figure~\ref{fig-acc} for the highlighted analyses. The second factor is the stop cross section (figure~\ref{fig-stop-cs}), again relative to the $t\bar t$ cross section. The final factor is the expected number of $t\bar t$ events, which we  take from the experimental analyses themselves, with one important modification -- we have normalized the $t\bar t$ numbers of events to the NNLO$_{\rm approx}$+NNLL cross sections quoted in appendix~\ref{sec-xsec}, which generally differ slightly (typically lower by $\sim 10\%$) from the values assumed in the original experimental analyses. Table~\ref{tab-events} shows the event yields for the case of a $150$~GeV stop. We estimate that the event yields are correct to the $\sim 10\%$ level (for more details, see appendix~\ref{sec-simulation}). This is sufficient for all practical purposes since this is anyway of the size of the theoretical uncertainty on the stop cross section.

\begin{figure}[t]
\begin{center}
\includegraphics[width=0.49\textwidth]{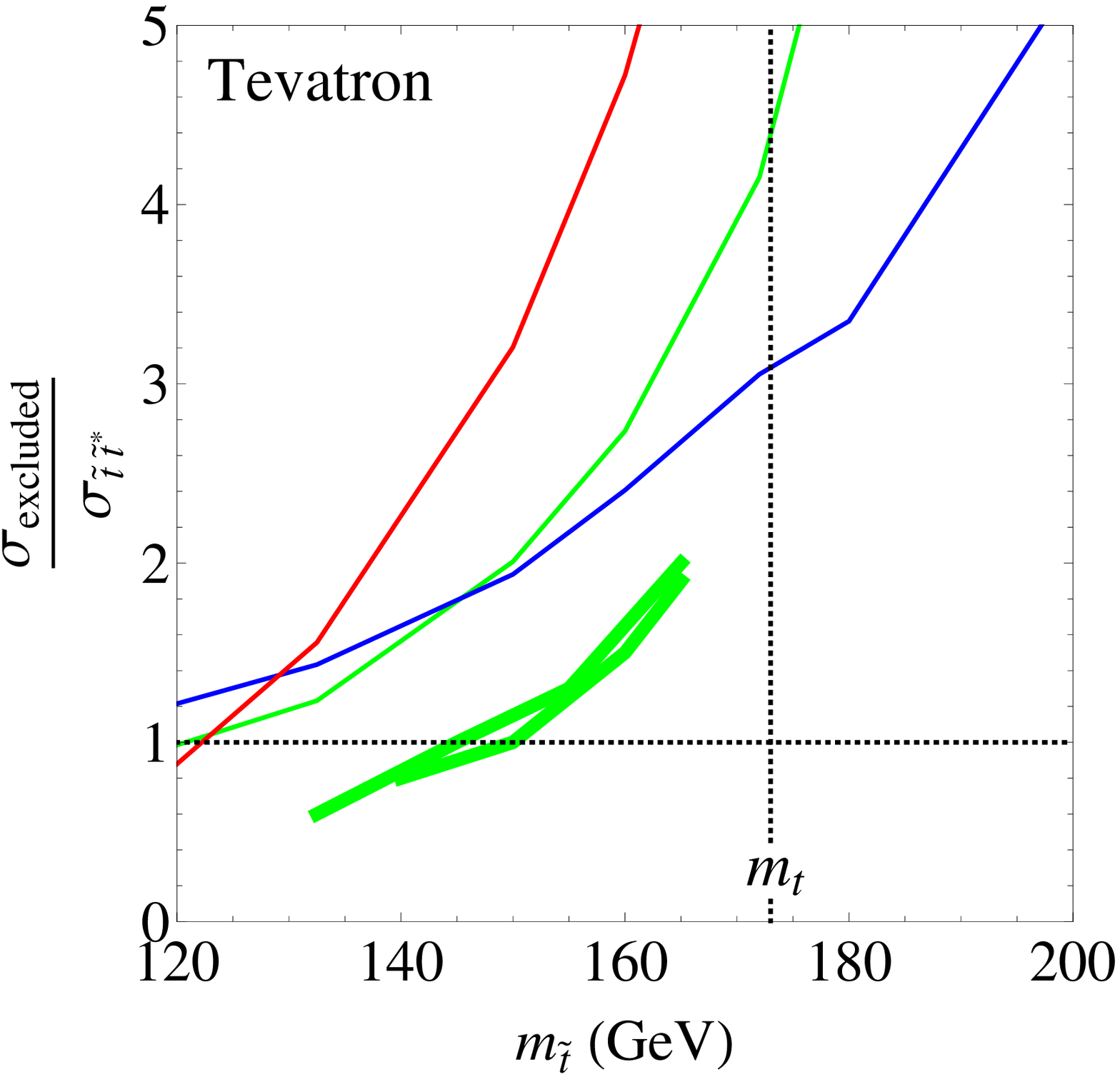}
\includegraphics[width=0.49\textwidth]{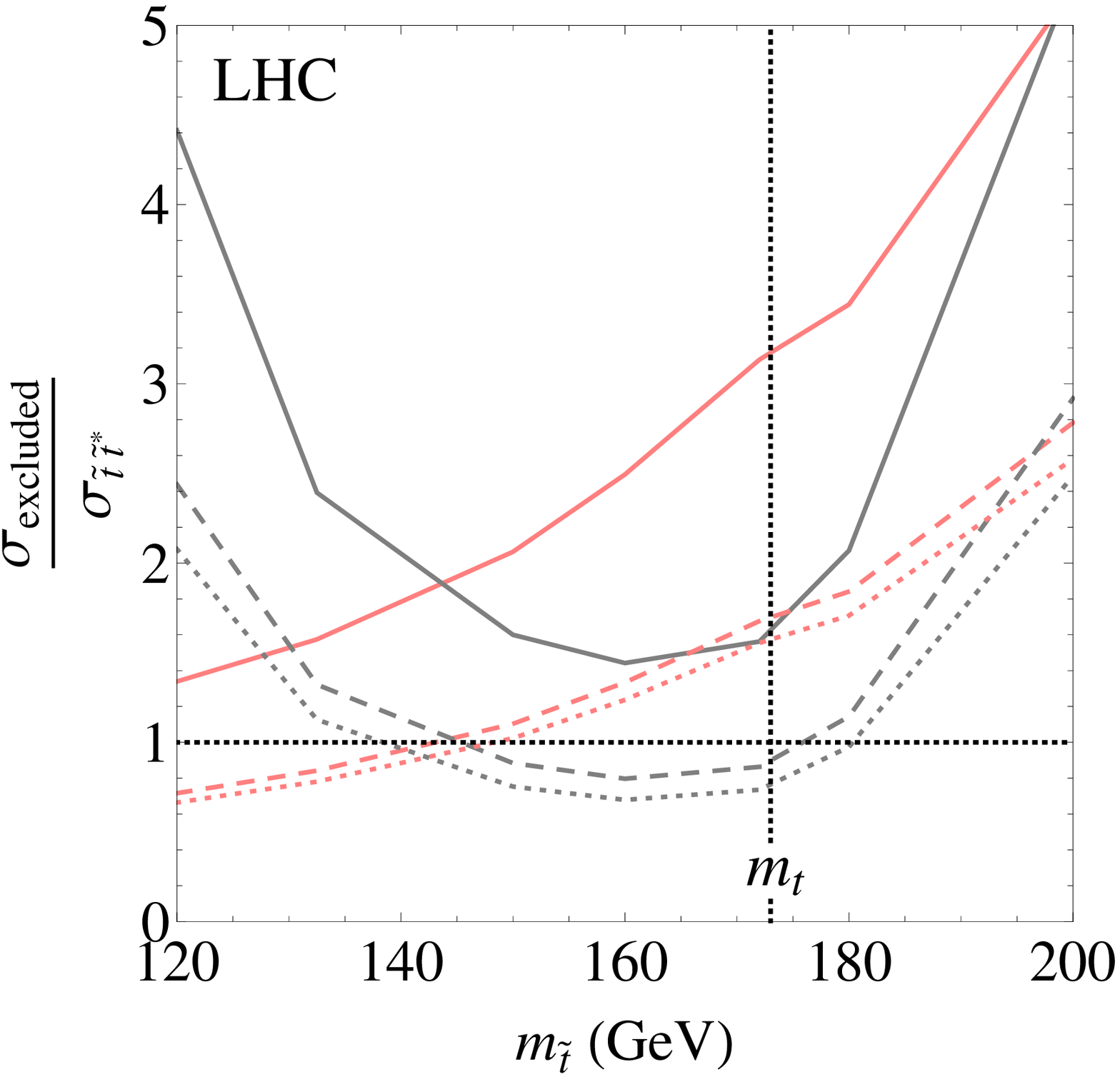}
\end{center}
\caption{$95\%$ CL excluded cross sections, relative to the theoretical stop cross section. On the left we present our limits from the Tevatron analyses: the pre-tag sample of the CDF $t\bar t$ cross section measurement in the dilepton channel~\cite{CDF-9890} (blue), the D0 stop search in the $e\mu$ channel (up to selection 1)~\cite{Abazov:2010xm} (red), and the $b$-tagged sample of the CDF stop search in the dilepton channel~\cite{Aaltonen:2009sf} (green). The two thick green lines are obtained from the stop mass reconstruction procedure of~\cite{Aaltonen:2009sf}. The right plot presents our limits from the LHC analyses: the pre-tag sample of the ATLAS $t\bar t$ cross section measurement in the dilepton channel~\cite{ATLAS-CONF-2011-034} (pink) and the ATLAS top partner search~\cite{ATLAS-CONF-2011-036} (gray), where the solid lines are the actual limits (from $35$~pb$^{-1}$ of data) while the dashed and dotted lines are approximate expected limits for $300$ and $3000$~pb$^{-1}$, respectively.}
\label{fig-95CLxsec}
\end{figure}

Finally, with the expected number of stop events in hand, we estimated the exclusion confidence levels (CL) for the stop NLSP as a function of its mass using the frequentist method~\cite{Conway:2000ju}. Our confidence levels include estimates of the systematic uncertainties based on the information available in the experimental papers~-- for more details see appendix~\ref{sec-err}. The $95\%$ CL excluded cross sections, relative to the theoretical stop cross section, are summarized in figure~\ref{fig-95CLxsec}.  For the LHC searches we also present the expected exclusion limits for $300$~pb$^{-1}$ and $3$~fb$^{-1}$ of data. We computed them by assuming the number of observed events for those luminosities to be the expected number of background events, and the systematic uncertainties (in $\%$) to remain the same. This is a conservative, somewhat pessimistic assumption since some of the systematics will improve with more data.

Figures~\ref{fig-acc} and~\ref{fig-95CLxsec} contain the main results of our paper. Some comments on these results are now in order.

By estimating the number of stop events via the ratios in (\ref{eq-nstop}) and the experimental prediction for $N_{t\bar t}$,  many of the systematic errors in our simulations should cancel out. This gives us some confidence in the accuracy of our results. Another important check is that the raw number of $t\bar t$ events predicted by our simulations generally agrees at the $\sim 30\%$ level with $N_{t\bar t}$ from the experimental references. (For more details see appendix~\ref{sec-simulation}.)  Note that in any event, the limits on the stop mass are fairly robust. Changing the acceptance by $\sim 10\%$ in either direction would not affect the stop mass limit by very much, given the rapid power-law dependence of the stop cross section as a function of its mass.

\begin{table}
$$\begin{array}{|c|c|c|c|c|c|}\hline
\mbox{Analysis}                                          & \st\st^\ast & \mbox{$t\bar t$ bg} & \mbox{total bg} &\,\mbox{data}\,\\\hline
\mbox{CDF $t\bar t$ cross section~\cite{CDF-9890} (pre-tag sample)}               &    21       & 156                 &  223\pm16       &    215      \\
\mbox{D0 stop search~\cite{Abazov:2010xm} (selection 1)}        &    46       & 183                 & 1174\pm73       &   1147      \\
\,\mbox{CDF stop search~\cite{Aaltonen:2009sf} ($b$-tagged sample)}   &    12.9     & 45.0                & 52.4\pm7.2      &     57      \\
\mbox{ATLAS $t\bar t$ cross section~\cite{ATLAS-CONF-2011-034} (pre-tag sample)}\,&\,  17.5   \,&\,74.7             \,&\,96.3\pm8.6   \,&    105      \\
\mbox{ATLAS top partner search~\cite{ATLAS-CONF-2011-036}}                  &     7.2     &  9.2                & 17.2\pm2.6      &     17      \\\hline
\end{array}$$
\caption{The expected numbers of events for a $150$~GeV stop NLSP, $t\bar t$ background, total background (and its systematic uncertainty, as discussed in appendix~\ref{sec-err}) and the observed number of events.}
\label{tab-events}
\end{table}

We find that for very light stops ($m_\st \sim 120$~GeV), the acceptance is affected significantly by the requirements regarding the number and $E_T$ of jets. Recall from figure~\ref{fig-pT-LHC} that the $b$ jets coming from light stop decays are very soft. In figure~\ref{fig-acc}, the acceptance is close to $1$ only for the D0 stop search (red curve) because that analysis does not impose a requirement on the number of jets. As a result, it gives the best cut-and-count exclusion limit for a $120$~GeV stop (see figure~\ref{fig-95CLxsec}). The acceptance is still relatively high for the $b$-tagged sample of the CDF stop search (green curve), which requires one jet of $15$~GeV and another of $12$~GeV. All the other analyses in figure~\ref{fig-acc} (and even more so some of the analyses from tables~\ref{tab-tt-analyses} and~\ref{tab-np-analyses} which are not included in the figure) have stricter jet $E_T$ requirements which result in much lower acceptances. This is true for both dilepton and lepton+jets analyses.

For heavier stops ($m_\st \gtrsim 150$~GeV) the $b$ jets become harder and are able to satisfy the selection requirements much more efficiently, but the limits become significantly weaker because of the smaller cross sections. The ATLAS top partner search in the lepton+jets channel has an exceptionally high acceptance relative to $t\bar t$, because it uses a cut on $m_T$ requiring $m_T > 120$~GeV. As shown in figure~\ref{fig-mT-LHC}, this eliminates the top background (where $m_T$ peaks sharply below $m_W$ since the MET is all from a single $W\to\ell\nu$ decay), while still retaining a decent fraction of the signal. This search is still unable to exclude any stop mass range at $95\%$ CL, but this is due to low statistics. We find that with several hundreds pb$^{-1}$ of data it will have a potential for $95\%$ CL exclusion for stop masses up to about $180$~GeV, as shown in figure~\ref{fig-95CLxsec}, but beyond that the analysis is limited by (our rather pessimistic extrapolation of) the systematic errors.

\subsection{Stop mass reconstruction\label{sec-mass-rec}}

\begin{figure}[t]
\begin{center}
\includegraphics[width=0.39\textwidth]{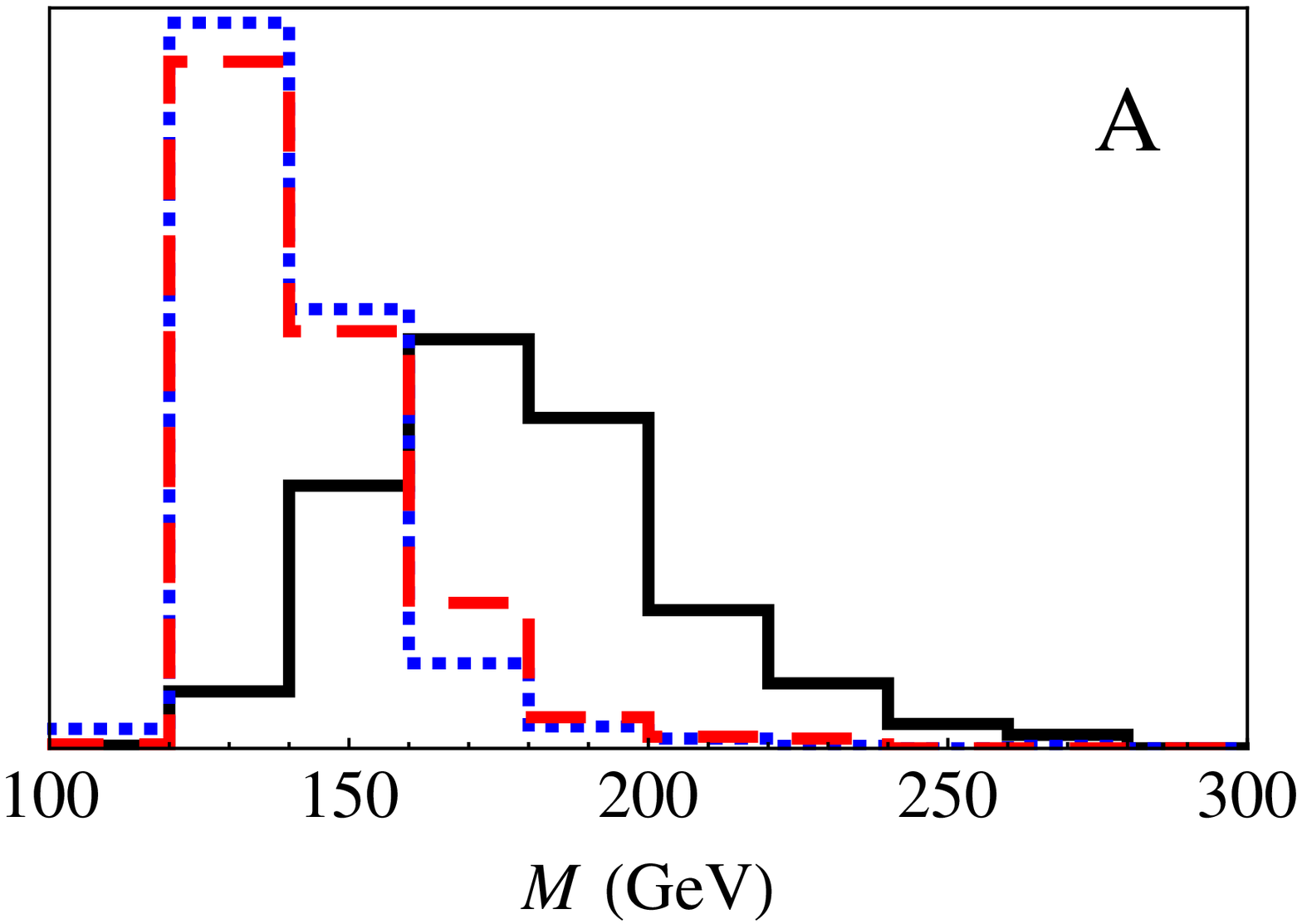}\q
\includegraphics[width=0.39\textwidth]{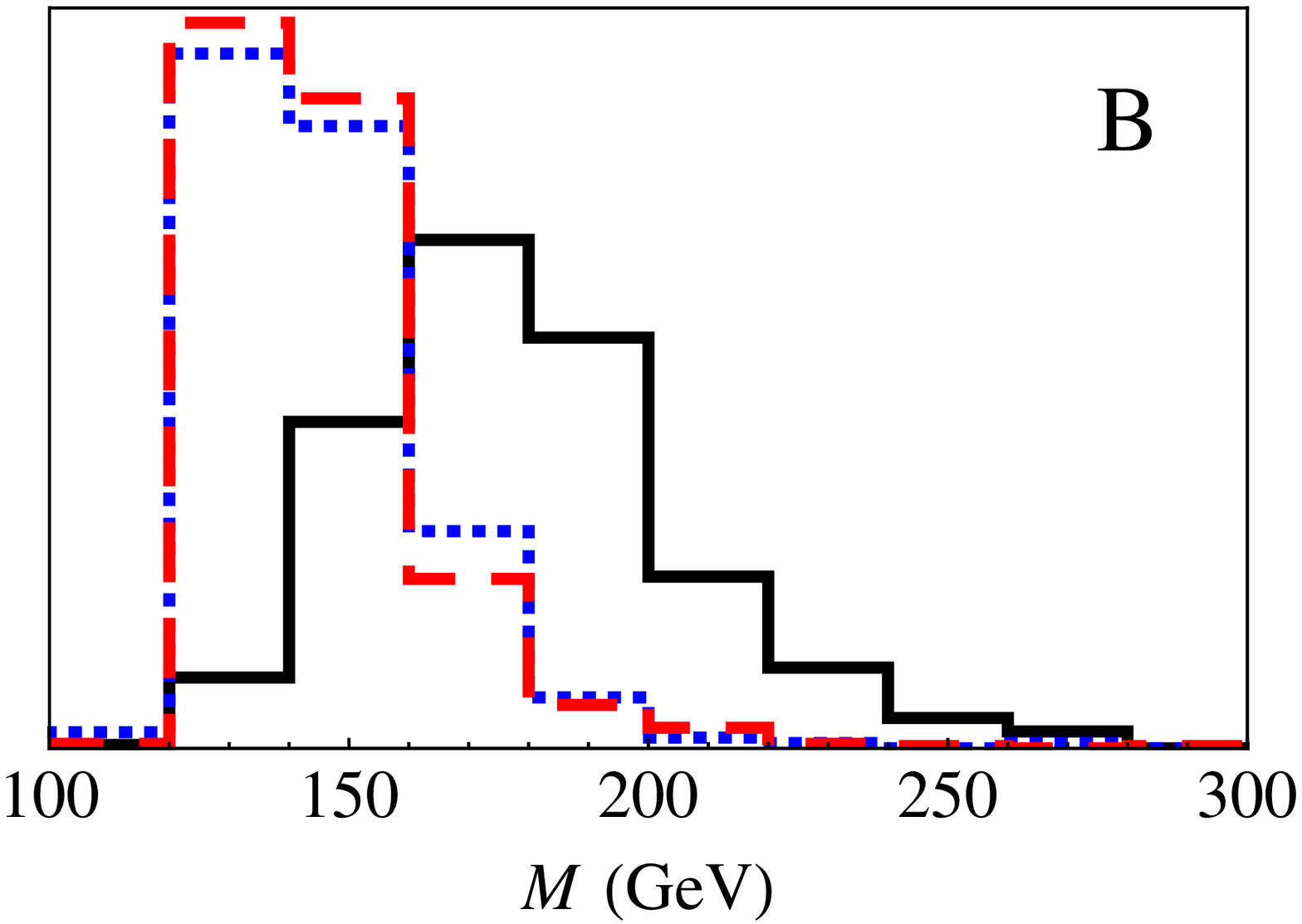}\\\vspace{5mm}
\includegraphics[width=0.39\textwidth]{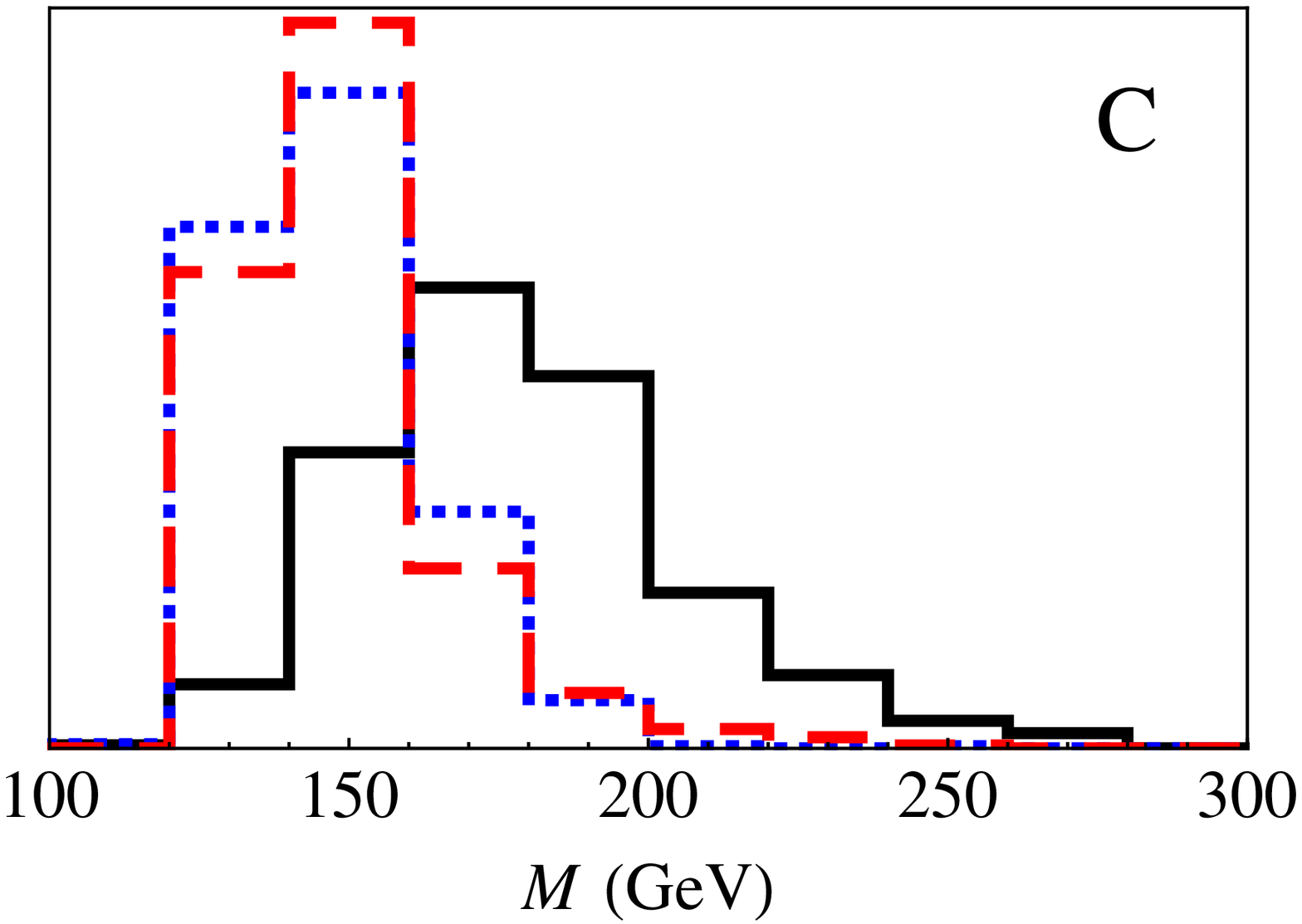}\q
\includegraphics[width=0.39\textwidth]{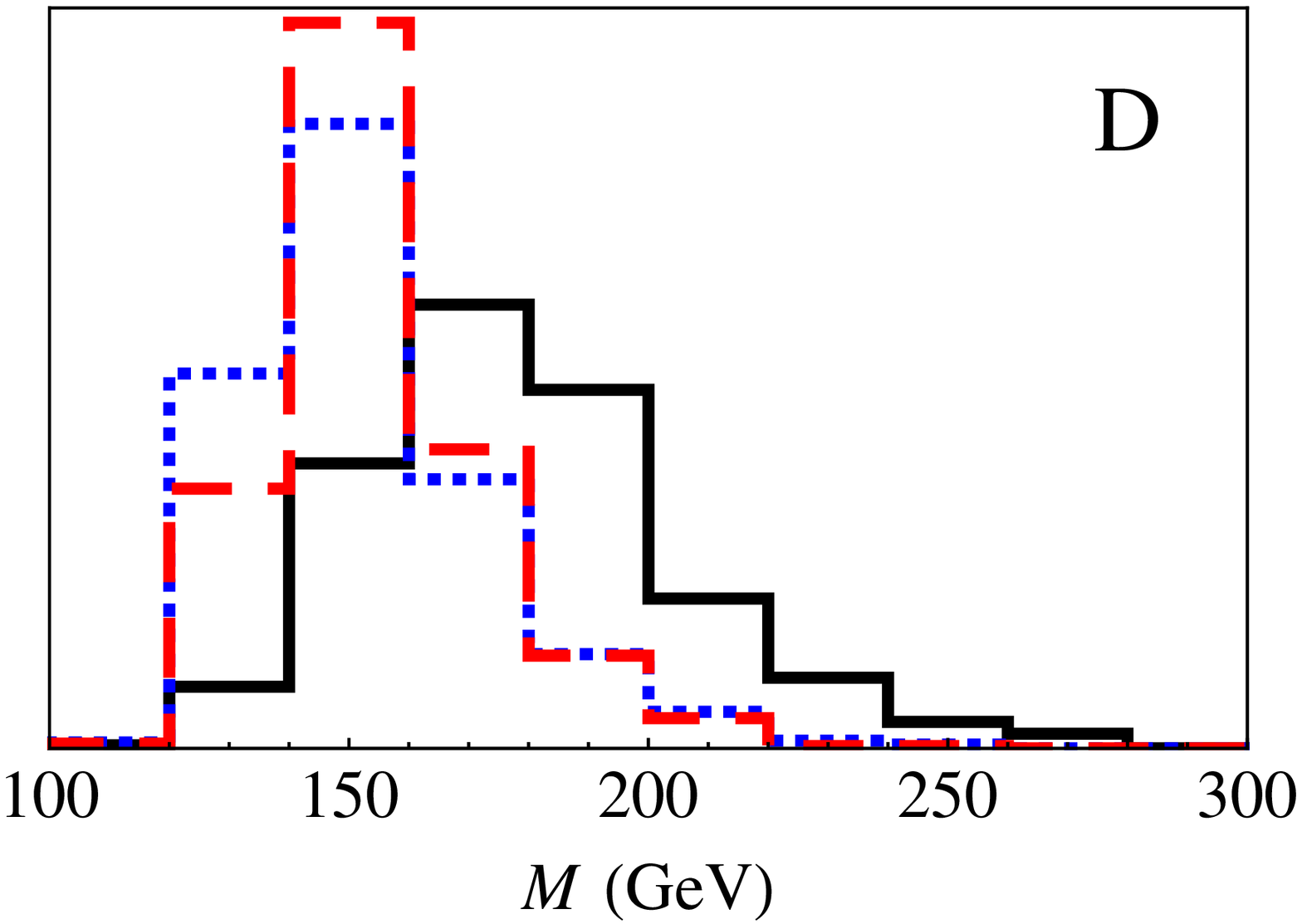}\\\vspace{5mm}
\includegraphics[width=0.39\textwidth]{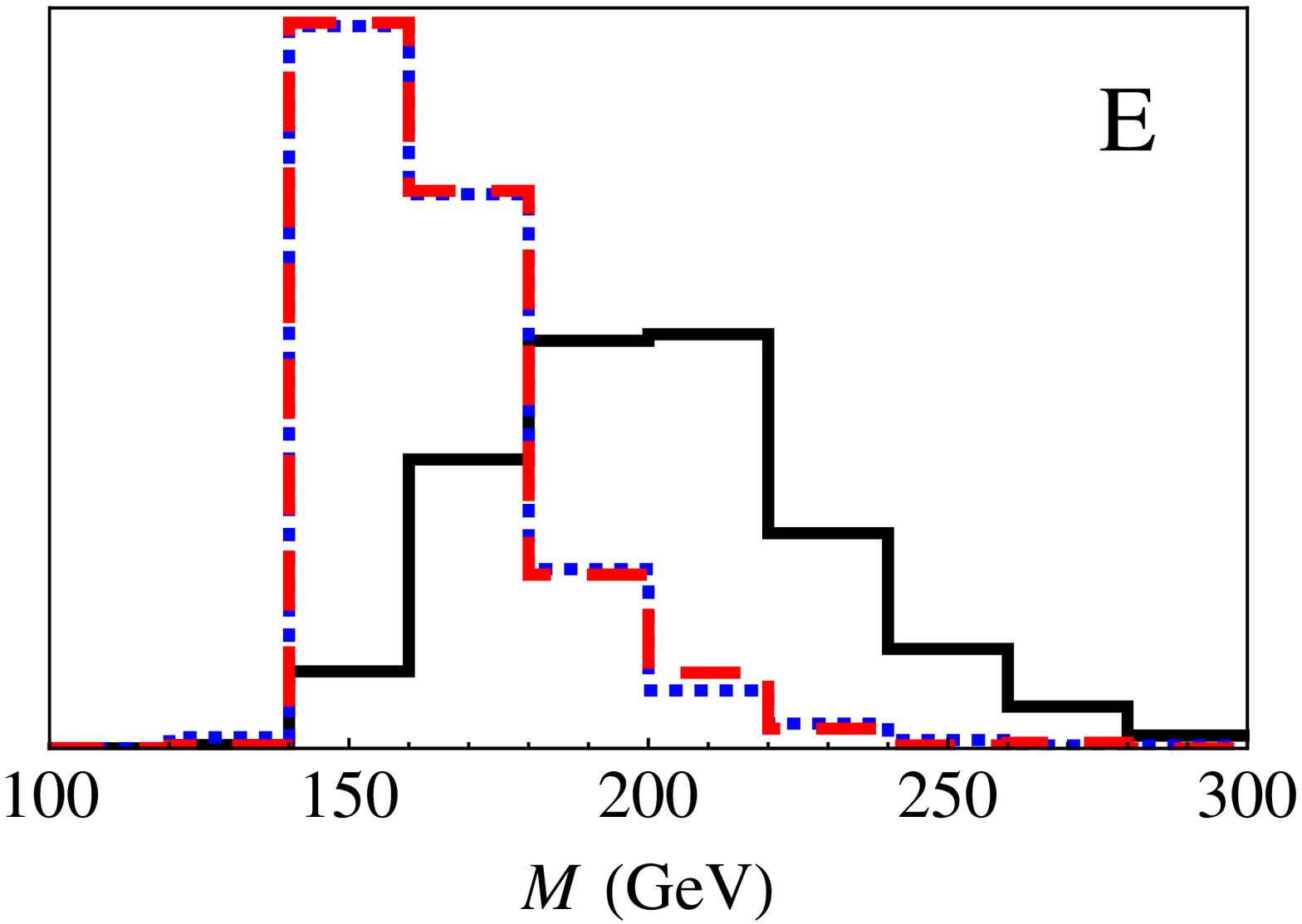}\q
\includegraphics[width=0.39\textwidth]{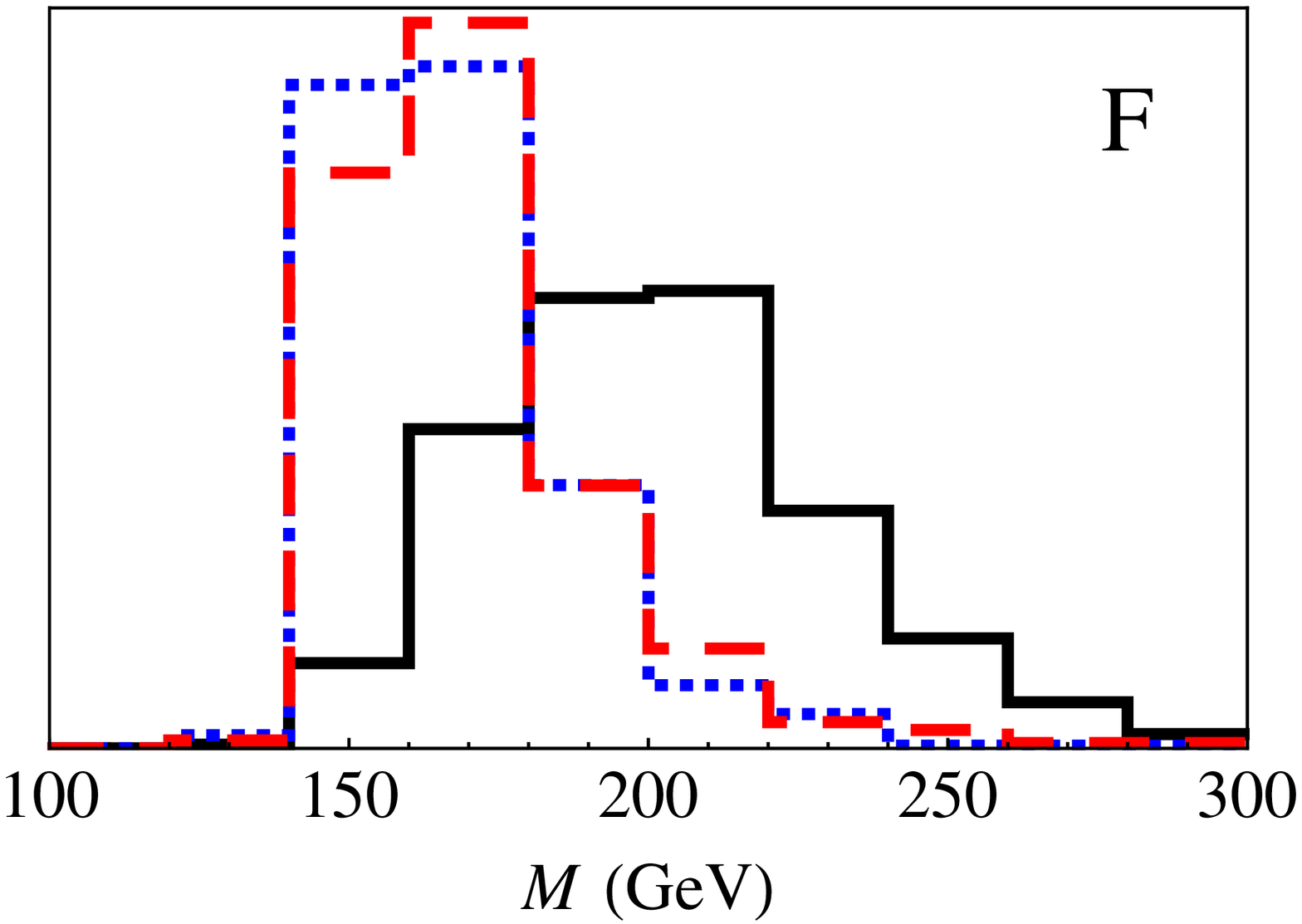}\\\vspace{5mm}
\includegraphics[width=0.39\textwidth]{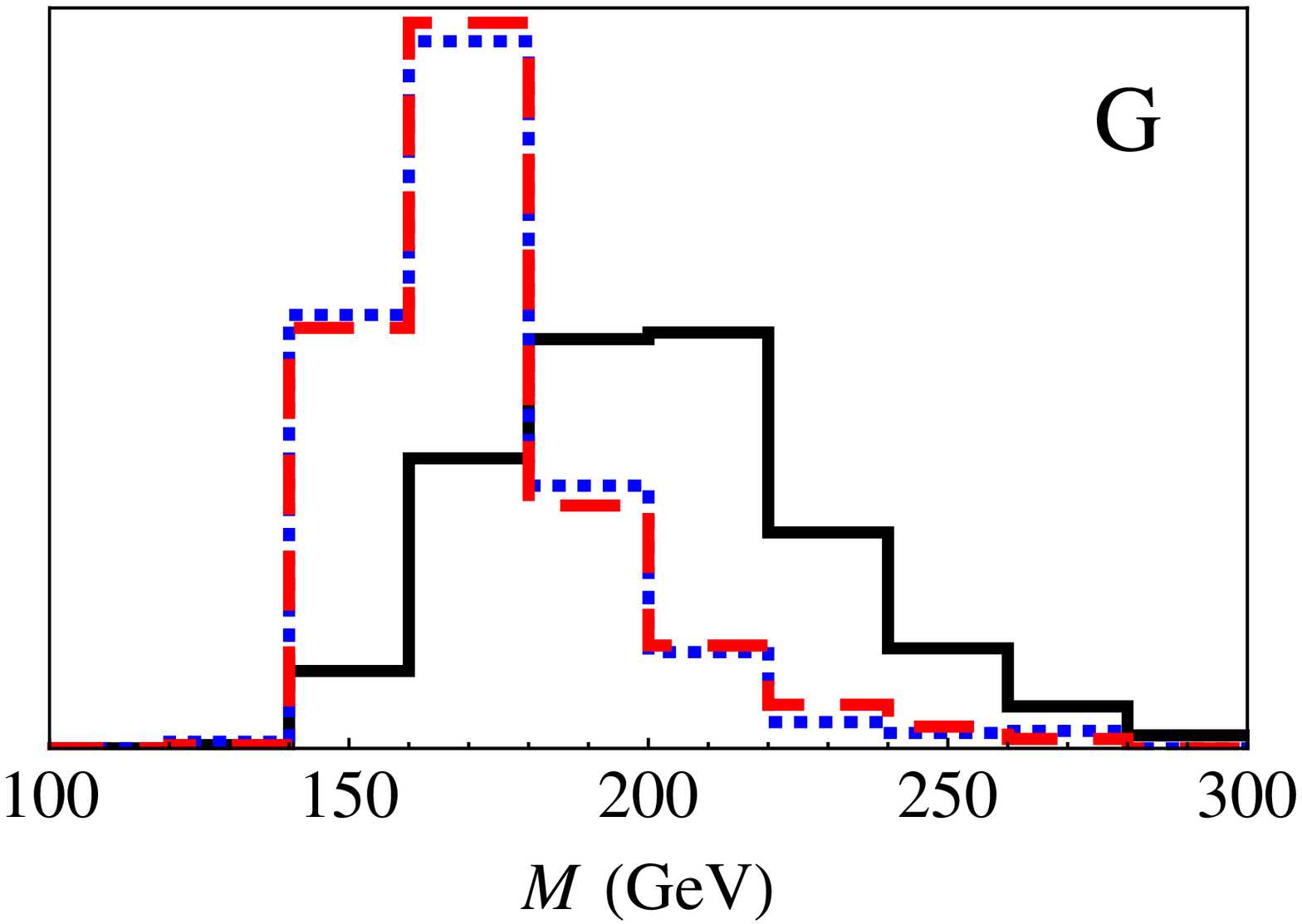}\q
\includegraphics[width=0.39\textwidth]{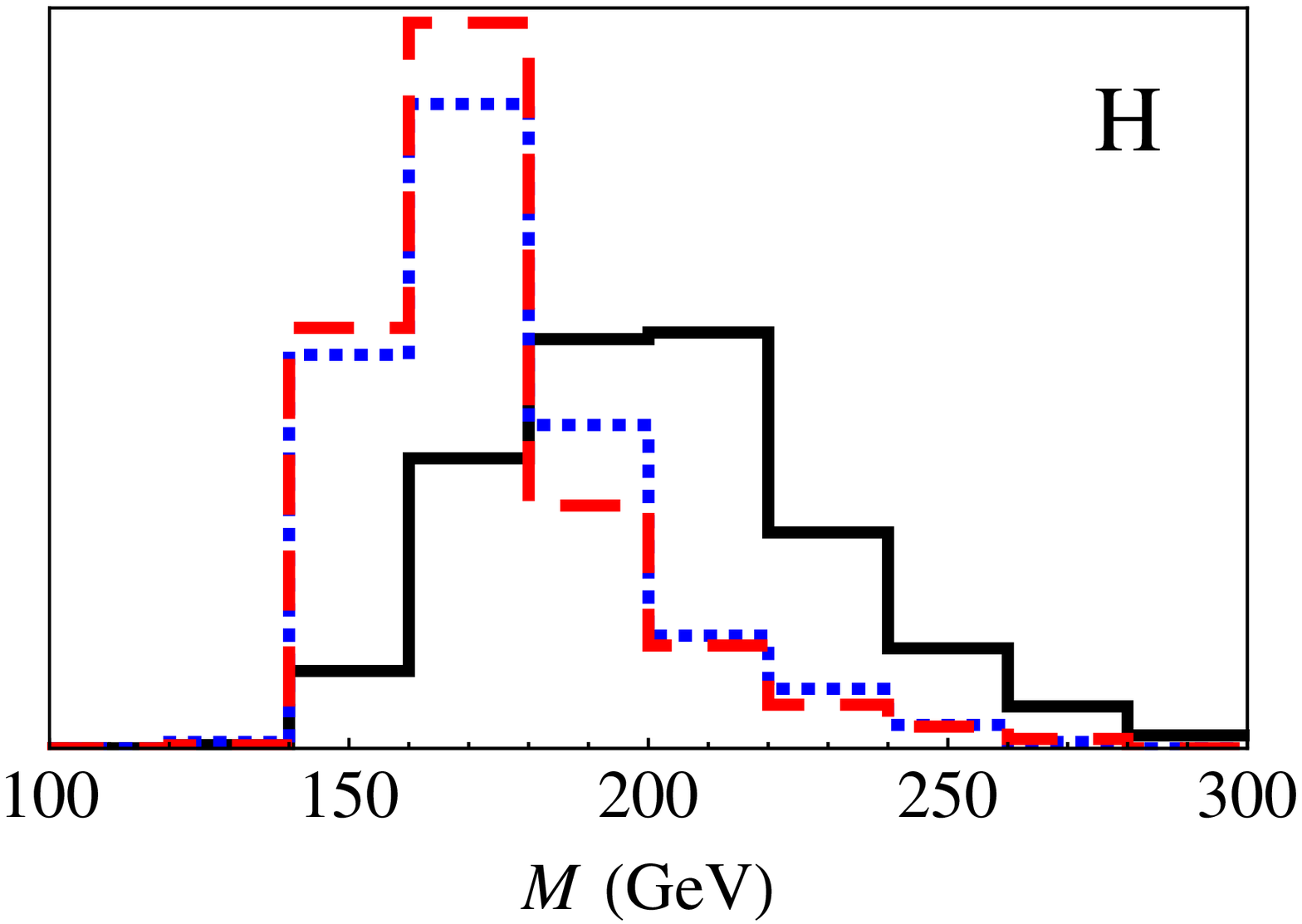}
\end{center}
\caption{Normalized distributions of the reconstructed stop mass obtained for the top (solid black), and the gravity-mediated (dashed red) and NLSP (dotted blue) stops from table~\ref{gravity-GMSB-correspondence}.}
\label{fig-mass-reconstuction}
\end{figure}

We have implemented the dilepton stop mass reconstruction algorithm of the CDF analysis~\cite{Aaltonen:2009sf,CDF-9439,Johnson:2010zza} (in a somewhat simplified way described in appendix~\ref{sec-stop-mass-algorithm}). Despite the fact that the logic of the algorithm uses the assumption that the stop decays as $\st \to b\tilde\chi_1^+ \to \ell^+\nu b\tilde\chi_1^0$, we find that surprisingly it works also for our stop NLSPs, in the sense of giving a reconstructed mass distribution that is much sharper than that of the top and centered at a different value. Our reproduction of the distribution from~\cite{Aaltonen:2009sf} (for $m_\st = 132.5$~GeV) is shown in figure~\ref{fig-mass-reconstuction}(A) (for the $b$-tagged channel). The same figure shows also the result we obtain for a stop NLSP of the same mass.

We can use the results of the CDF study for setting limits on stop NLSPs by finding pairs of NLSP and gravity-mediated stops with similar mass distribution (which would sometimes require them to have non-equal true masses) and same number of events (which can be obtained by tuning the dilepton branching ratio of the gravity-mediated stop). Note that by taking this approach we bypass the need to reproduce the analysis of the systematic uncertainties. Also, the imperfections in our simulation of the stop NLSP events and the mass reconstruction algorithm are likely to be canceled to a large extent by similar imperfections in our simulation of the corresponding gravity-mediated stops.

\begin{table}[t]
$$\begin{array}{|c||c||c|c|c|c|}\hline
\,\mbox{case} &\, \mbox{NLSP} \,& \multicolumn{4}{c|}{\mbox{gravity-mediated}} \\\hline
& m_\st & m_\st & m_{\tilde\chi_1^\pm} & m_{\tilde\chi_1^0} & \mbox{BR} \\\hline
\mbox{A} & 132.5 &\, 132.5 \,&\, 105.8 \,&\, 47.6 \,&\, 0.17 \,\\
\mbox{B} & 145   &   137     &   105.8   &   47.6   &   0.13   \\
\mbox{C} & 155   &   140     &   105.8   &   47.6   &   0.11   \\
\mbox{D} & 165   &   150     &   105.8   &   47.6   &   0.10   \\\hline
\end{array}\q
\begin{array}{|c||c||c|c|c|c|}\hline
\,\mbox{case} &\, \mbox{NLSP} \,& \multicolumn{4}{c|}{\mbox{gravity-mediated}} \\\hline
& m_\st & m_\st & m_{\tilde\chi_1^\pm} & m_{\tilde\chi_1^0} & \mbox{BR} \\\hline
\mbox{E} & 140   &\, 150   \,&\, 125.8 \,&\, 58.8 \,&\, 0.25 \,\\
\mbox{F} & 150   &   155     &   125.8   &   58.8   &   0.20   \\
\mbox{G} & 160   &   160     &   125.8   &   58.8   &   0.16   \\
\mbox{H} & 165   &   160     &   125.8   &   58.8   &   0.13   \\\hline
\end{array}$$
\caption{Pairs of NLSP--gravity-mediated stops which have equal numbers of events passing the selection of the $b$-tagged channel of~\cite{Aaltonen:2009sf} and similar reconstructed mass distributions (figure~\ref{fig-mass-reconstuction}). The first table corresponds to figure 2(a) of~\cite{Aaltonen:2009sf} which assumes chargino mass of $105.8$~GeV and the second table to figure 2(b) with chargino mass of $125.8$~GeV.}
\label{gravity-GMSB-correspondence}
\end{table}

Table~\ref{gravity-GMSB-correspondence} shows pairs of NLSP and gravity-mediated stops that have same event counts and similar reconstructed mass plots (in the $b$-tagged channel) which are shown in figure~\ref{fig-mass-reconstuction}. With these pairs in hand, we can approximately read off the corresponding $95\%$ CL exclusion limits from figure~2 of~\cite{Aaltonen:2009sf}. We include the results in figure~\ref{fig-95CLxsec} (thick green lines, corresponding to the two chargino mass hypotheses of~\cite{Aaltonen:2009sf}). The conclusion is that stop NLSP is excluded for $m_\st \lesssim 150$~GeV.

Evidently, the more sophisticated approach of the CDF stop search sets the best limit on stop NLSPs. This illustrates the  power of using more discriminating variables in searching for new physics in the $t\bar t$ sample. We are optimistic that with more work the limit could be further improved with existing data. For instance, it would be interesting to see what constraints the existing D0 stop search~\cite{Abazov:2010xm} can set when going beyond the simple cut-and-count portion (``selection 1'') that we have considered here.

\section{Other types of measurements\label{sec-other-measurements}}
\setcounter{equation}{0}

In this section we discuss several additional methods that may be relevant for future searches for stop NLSPs.

\subsection{$m_{\ell b}$ and $b$-jet $p_T$} \label{sec-mlb}

As has been pointed out already in~\cite{Chou:1999zb}, the stop NLSP has a distinct distribution of the invariant mass $m_{\ell b}$ which can be useful for reducing the $t\bar t$ background. We show this distribution in figure~\ref{fig-mlb-LHC}. While there exists an ambiguity in pairing each lepton with the $b$ jet that came from the same top or stop, we see that even incorrectly paired cases happen to contribute in a similar way to the differences between the distributions. This happens because the lower $m_{\ell b}$ values for the stops can be attributed to a large extent to the lower momenta of the $b$ quarks. Unfortunately, it seems that raw data for the $m_{\ell b}$ distribution of $t\bar t$ samples at the Tevatron have not been published since the early measurements with $\sim 700$~pb$^{-1}$~\cite{Abulencia:2006iy}. Such data from the Tevatron or the LHC may be able to strongly constrain the stop.

It is also possible that just the analysis of the $b$-jet $p_T$ distribution will have a comparable power. As discussed in section~\ref{sec-kin-dist} (see figure~\ref{fig-pT-LHC}), the $b$-jets coming from stop NLSP decays are generally much softer than those from top decays. If some way could be devised to take advantage of this separation, this could become a useful way of distinguishing stop events from top events. Some potential limiting factors for such an analysis might be: that only jets with $p_T \gtrsim 12$~GeV can be properly reconstructed; that without $b$ tagging ISR jets may contribute as well; that $b$ tagging efficiency decreases with $p_T$; and that additional backgrounds become important once low-$p_T$ jets are allowed.

\subsection{Displaced decays}

As we mentioned in section~\ref{sec-stop-NLSP}, the stop NLSP can naturally be long lived. (Another situation in which the stop can be long-lived due to a suppressed coupling is the right-handed sneutrino LSP scenario~\cite{deGouvea:2006wd}.) If the stop decays between $\sim 100$~$\mu$m and $\sim 0.5$~m away from the interaction point in the plane transverse to the beam (the precise range depends on the particular experiment), the tracks of the charged particles produced from its decay products can be identified as emerging from a displaced vertex. For distances above $\sim 1$~cm, where the background of displaced vertices coming from heavy flavor production becomes small, the displaced vertices of the stop can even make its signal easier to separate from the $t\bar t$ and other standard model backgrounds. Triggering on such events is easy thanks to the presence of leptons (from the leptonic decays of the $b$ or the $W$). Reconstructing the displaced vertices is also simple because of the presence of a $b$ jet and a lepton or two additional jets, each jet typically containing multiple charged particles.

Even without using the presence of displaced vertices, such events may be included in the samples of prompt searches like those we studied in this paper and our analysis will apply. But this depends on the details of the logic used in the event selection procedure of each experiment. For example, such events may not make it into the sample or may be analyzed incorrectly if some part of the procedure uses the assumption that all the relevant tracks in the event emerge from the vicinity of the primary vertex.\footnote{For example, the D0 stop search~\cite{Abazov:2010xm} requires the distance of closest approach between the muon track and the primary vertex to be $< 0.02$~cm if the track
includes hits in the silicon tracker or $< 0.2$~cm otherwise.}

So far, there have not been any dedicated searches for stop decays at displaced vertices. However, many of the ideas that would be useful for such a search can be borrowed from searches that addressed displaced decays of certain other particles. In particular, the D0 search~\cite{Abazov:2009ik} for pair produced long-lived neutral particles that each decay into $b\bar b$ within the tracker (motivated by the hidden valley scenario) seems relevant. Methods for studying such a scenario have been developed also at ATLAS~\cite{ATL-PHYS-PUB-2009-082}, who in addition considered the possibility that the decays occur within the calorimeters. The large missing energy present in the stop NLSP scenario, as well as the fact that the stops are somewhat heavier than the particles considered by D0 and ATLAS, may be helpful for further eliminating backgrounds. The D0 search (with $3.6$~fb$^{-1}$) has excluded signals with cross sections of the order of $10$~pb. It is therefore plausible that such a method will be efficient for studying the displaced stop NLSP scenario.

Furthermore, unlike in the scenario described in the previous paragraph where the particles are neutral, the stops will hadronize with light quarks into either neutral or charged particles~\cite{Sarid:1999zx}. In the latter case, they will have tracks characterized by an anomalously large rate of energy loss through ionization ($dE/dx$) which can be measured if they traverse a sufficiently long distance within the tracking volume. In fact, this kind of measurement has been already used by CMS~\cite{Khachatryan:2011ts} (with $3.1$~pb$^{-1}$) and ATLAS~\cite{Aad:2011yf} (with $34$~pb$^{-1}$), although without any results relevant to displaced stops. The CMS study did not have the reach to exclude stops of any mass. It should also be noted that the study assumed the stops to be stable throughout the calorimeters and computed the trigger efficiencies accordingly. If the stop decays earlier, the signal should be simulated differently. For example, the missing energy in the scenario considered by CMS is carried by the stops themselves (which similarly to muons do not deposit much of their energy in the calorimeters) while in our case it would be carried by the gravitinos and sometimes the neutrinos. The ATLAS study required the tracks to match to either a reconstructed ``muon'' in the muon calorimeter or to a cluster in the tile calorimeter. As a result, their analysis does not have acceptance for stops that decay before reaching those.

\subsection{Bound states}

An interesting feature of the stop NLSP scenario is that the near-threshold $\st\st^\ast$ bound state, the stoponium, is guaranteed to decay by annihilation since its annihilation rate (into $gg$) is
\be
\Gamma_{\rm annih} \approx \frac{32}{81}\alpha_s^5 m_\st \sim \l(10^{-13}\mbox{ m}\r)^{-1}
\ee
where we evaluated the decay length for $m_\st \sim m_t$ (compare with figure~\ref{fig-sqrtF}). Such a stoponium will be observable at the LHC within a few years as a narrow diphoton resonance from
\be
(\st\st^\ast) \to \gamma\gamma
\ee
at invariant mass of slightly below $2m_\st$ (for recent discussions, see~\cite{Martin:2008sv,Kahawala:2011pc}). The observation of this signal will allow a precise measurement of the stop mass. The size of the signal and the angular distribution will also help confirm the identity and the properties of the stop.

\subsection{Flavor-violating decays and same-sign dilepton signals}

If the MSSM has flavor violation beyond that of the Standard Model (which is not the case in gauge mediation), the process
\be
\st \to c\,\gr
\label{stop-decay-FV}
\ee
may become significant and even dominate over~(\ref{stop-decay})~\cite{Sarid:1999zx}. While we are unaware of any existing searches of (\ref{stop-decay-FV}), a similar process
\be
\st\to c\,\tilde\chi_1^0
\ee
(with a massive neutralino $\tilde\chi_1^0$) has been considered by D0~\cite{Abazov:2008rc} and CDF~\cite{CDF-9834}. The CDF search was able to exclude stops up to $180$~GeV in the case that this process dominates and $m_\st - m_{\tilde\chi_1^0} \gtrsim 50$~GeV. The limit would probably be even stronger in the case of (\ref{stop-decay-FV}) -- since the gravitino is massless, the $c$-jet $p_T$ and the missing energy would both be larger. Since we are mostly interested in light stops ($m_\st \lesssim m_t$), the contribution of (\ref{stop-decay-FV}) can be considered small.

However, even a relatively small extra flavor violation, for which (\ref{stop-decay}) still dominates, may lead to the conversion of the stops into antistops when they bind with light quarks into neutral mesinos before decaying~\cite{Sarid:1999zx}. This would lead to same-sign dilepton events with jets and missing energy which have very little background. The non-observation of such events~\cite{CDF-10464,Chatrchyan:2011wb} will provide support to the gauge mediation scenario if a light stop is observed through (\ref{stop-decay}).

\section{Conclusions\label{sec-conclusions}}
\setcounter{equation}{0}

We hope we have convinced the reader that there is a viable possibility for the stop to be light, and hiding in the $t\bar t$ sample. On the other hand, we have shown that some existing and ongoing Tevatron and LHC searches do have sensitivity to this scenario. Now is therefore the right time to study it and perhaps discover supersymmetry in our backyard while the LHC is pushing the other squarks and the gluino to higher and higher masses.

Besides determining the fate of the light stop NLSP, the relevant searches would also be useful for gaining more confidence in the purity of the top sample or maybe discovering a different new physics contribution within it. This seems especially important in view of the anomalous forward-backward asymmetry observed at the Tevatron.

Our limit $m_\st \gtrsim 150$~GeV was obtained by analyzing measurements that were designed for other purposes. We have also seen that even without any further optimization, existing LHC searches should have sensitivity to the entire range of $m_{\st}<m_t$ with just 300~pb$^{-1}$. These results can almost certainly be improved by performing more dedicated searches for stop NLSP at the Tevatron and LHC. For example, our simulation of the simple cuts stage of the D0 search~\cite{Abazov:2010xm} has already allowed us to set the limit $m_\st \gtrsim 123$~GeV, but without having access to the data we were unable to utilize the full strength of the D0 approach which used distributions of a set of powerful discriminating variables optimized for a different stop scenario. Adapting this kind of search for the stop NLSP scenario may improve our $m_\st \gtrsim 150$~GeV limit. Even more simply, the signal-to-background ratios for many  analyses will benefit from allowing for softer jets and/or using more discriminating variables such as $m_T$ or $m_{\ell b}$ to reduce backgrounds. Clearly, dedicated searches for the stop NLSP scenario would be very beneficial.

In this paper, we have focused on prompt decays of stop NLSPs. This has allowed us to cast our results in a 1D parameter space consisting solely of the stop mass. However, as we have discussed, it is natural for the stop to be long-lived. Therefore, a larger 2D parameter space is relevant, consisting of the stop mass and lifetime. Most of this parameter space is unexplored territory. For instance, while the case of a detector-stable stop has been analyzed in~\cite{Aad:2011yf}, we are not aware of any searches which constrain stops with displaced decays inside the detector. Furthermore, even in searches optimized for either of the two extreme cases (prompt and detector-stable), it is not known how the acceptance degrades as one transitions to the intermediate lifetime regime. Ideally, the results of future searches for stop NLSP will be cast in the mass-lifetime plane, yielding a complete characterization of this simple, minimal, and well-motivated new physics scenario.

\section*{Acknowledgments}

We are grateful to Robin Erbacher, Tobias Golling, Eva Halkiadakis, William Johnson, Qiuguang Liu, Dennis Mackin, Fabrizio Margaroli, Chang-Seong Moon, Stephen Mrenna, Michael Peskin, Yoshitaro Takaesu, Jian Tang and Scott Thomas for helpful correspondence and discussions. The research of YK is supported in part by DOE grant DE-FG02-96ER40959. The research of DS is supported in part by a DOE Early Career Award.

\appendix

\section{Matrix element for stop decay\label{sec-ME}}
\setcounter{equation}{0}

The gravitino field $\psi_\mu$ couples to the supercurrent $S^\mu$ as (see, e.g.,~\cite{Moroi:1995fs,Pradler:2007ne,Luo:2010he})
\bea
\cL_{\rm int} &=& -\frac{i}{\sqrt2\,M_P}\bar\psi_\mu S^\mu \nn\\
&=& -\frac{i}{\sqrt 2\,M_P}\sum_{(\phi,\,\chi)}\l(D_\nu \phi\r)^\dagger\bar\psi_\mu\gamma^\nu\gamma^\mu\,\chi + \mbox{h.c.}
- \frac{i}{8M_P}\bar\psi_\mu \l[\gamma^\rho,\gamma^\sigma\r] \gamma^\mu\lambda^a F_{\rho\sigma}^a
\label{full-ME}
\eea
where $(\phi,\,\chi)$ are the scalar and the spinor components of a chiral multiplet and $M_P \equiv 1/\sqrt{8\pi G} \simeq 2.435\times 10^{18}$~GeV is the reduced Planck mass. The interactions relevant to us involve the $SU(2)$ singlet
\be
\phi = \st_R\,,\qq \chi = P_R t
\ee
and the $SU(2)$ doublet
\be
\phi = \mat \st_L \\ \tilde b_L \rix , \qq
\chi = \mat P_L t \\ P_L b \rix
\ee
Using $c_\st \equiv \cos\theta_\st$, $s_\st \equiv \sin\theta_\st$ to characterize the mixing of the stops, the vertices relevant to the lighter stop eigenstate (which would be completely left-handed for $s_\st = 0$) are
\be
\cL_{\st t\gr} = -\frac{i}{\sqrt 2\,M_P}\pd_\nu \st^\ast\,\bar\psi_\mu\gamma^\nu\gamma^\mu \l(c_\st P_L + s_\st P_R\r)t + \mbox{h.c.}
\ee
\be
\cL_{\st Wb\gr} = -\frac{1}{2M_P}g\, W_\nu^+\st^\ast\,\bar\psi_\mu\gamma^\nu\gamma^\mu\,c_\st P_L b + \mbox{h.c.}
\ee
where we used $D_\mu \equiv \pd_\mu + i g W_\mu^a T^a + \ldots\,$, where $g = e/\sin\theta_W$.

At energies $\gg m_\gr$, the gravitino $\psi_\mu$ can be replaced by the goldstino $\psi$ via~\cite{Moroi:1995fs,Kim:1997iwa,Luo:2010he}
\be
\psi_\mu \to -\sqrt\frac{2}{3}\, \frac{\pd_\mu\psi}{m_\gr}
\label{gravitino2goldstino}
\ee
Using the gravitino mass expression $m_\gr = F/\sqrt3\,M_P$ where $\sqrt F$ is the SUSY breaking scale, we get
\be
\cL_{\st t\gr} = \frac{i}{F}\,\pd_\nu \st^\ast\,\pd_\mu\bar\psi\gamma^\nu\gamma^\mu \l(c_\st P_L + s_\st P_R\r)t + \mbox{h.c.}
\label{st-t-G}
\ee
\be
\cL_{\st Wb\gr} = \frac{1}{\sqrt2\,F}\, g\, W_\nu^+\st^\ast\,\pd_\mu\bar\psi\gamma^\nu\gamma^\mu\,c_\st P_L b + \mbox{h.c.}
\ee
We can use
$\{\gamma^\mu,\gamma^\nu\} = 2g^{\mu\nu}$ and $\slash\pd\psi = 0$ (for an on-shell goldstino) to make the replacement
\be
\pd_\mu\bar\psi\gamma^\nu\gamma^\mu
= 2\,\pd^\nu\bar\psi - \pd_\mu\bar\psi\gamma^\mu\gamma^\nu
\to 2\,\pd^\nu\bar\psi
\ee
which gives (as in eq.~(51) of~\cite{Luo:2010he})
\be
\cL_{\st t\gr} = \frac{2i}{F}\l[\pd_\mu\st^\ast\, \pd^\mu\bar\psi\l(c_\st P_L + s_\st P_R\r) t - \pd_\mu\st\, \bar t\l(c_\st P_R + s_\st P_L\r)\pd^\mu\psi\r]
\label{st-t-G-onshell}
\ee
and (as in eq.~(56) of~\cite{Luo:2010he})
\be
\cL_{\st W b \gr} = \frac{\sqrt2}{F}\, g\, c_\st \l(W_\mu^+\,\st^\ast\,\pd^\mu\bar\psi P_L b + W_\mu^-\,\st\,\bar b P_R\,\pd^\mu\psi\r)
\label{contact}
\ee
Using (\ref{st-t-G}), (\ref{contact}), and the standard model interaction
\be
\cL_{tWb} = -\frac{g}{\sqrt2} \l(W_\mu^+\, \bar t\gamma^\mu P_L b + W_\mu^-\, \bar b\gamma^\mu P_L t\r)
\label{top-decay}
\ee
we obtain the matrix element (\ref{ME}) quoted in the text.

\section{Details relevant to exclusion limits}
\setcounter{equation}{0}

\subsection{$t\bar t$ and $\st\st^\ast$ production cross sections\label{sec-xsec}}

NNLO$_{\rm approx}$+NNLL calculations give the $t\bar t$ production cross section at the Tevatron as $\sigma_{t\bar t} = 6.30 \pm 0.19^{+0.31}_{-0.23}$~pb~\cite{Ahrens:2010zv} or $7.08^{+0.20+0.36}_{-0.24-0.27}$~pb~\cite{Kidonakis:2010dk}. As a compromise between the two sources we will assume $\sigma_{t\bar t} = 6.7\mbox{ pb} \pm 6\%$. For $7$~TeV LHC, the result is $\sigma_{t\bar t} = 149 \pm 7 \pm 8$~pb~\cite{Ahrens:2010zv} or $163^{+7+9}_{-5-9}$~pb~\cite{Kidonakis:2010dk}. We will use $\sigma_{t\bar t} = 156\mbox{ pb} \pm 7\%$.

NLO+NLL predictions for $\st\st^\ast$ production cross sections are available in~\cite{Beenakker:2010nq}. We model their results for the Tevatron by using the leading-order expressions (see, e.g.,~\cite{Beenakker:1997ut}) with MSTW 2008 NLO PDFs multiplied by a $K$-factor whose value and uncertainty vary linearly between $1.58 \pm 13\%$ and $1.48 \pm 9\%$ as the stop mass varies between $100$ and $200$~GeV. For the $7$~TeV LHC we model the results similarly with a $K$-factor of $1.73 \pm 13\%$. The cross sections are plotted in figure~\ref{fig-stop-cs}.

\subsection{Systematic uncertainties\label{sec-err}}

This section describes the systematic uncertainties we assume for the various cut-and-count experiments that we analyze.

For backgrounds other than $t\bar t$ we always use the uncertainties computed by the experimental groups. For $t\bar t$, we also use the result obtained by the experimentalists whenever available. In cases where it is not given explicitly we compute it as follows. We include a $6\%$ (Tevatron) or $7\%$ (LHC) uncertainty in the theoretical cross section and a $6\%$ (Tevatron) or $3.5\%$ (LHC) luminosity uncertainty. For the CDF analyses of~\cite{CDF-9890} (or~\cite{CDF-10163}) we also include $5.0\%$ (or $4.8\%$) acceptance uncertainty (due to the jet energy scale etc.), and for the $b$-tagged samples also $3.5\%$ ($4.1\%$) uncertainty from $b$ tagging efficiency. For the ATLAS analysis~\cite{ATLAS-CONF-2011-035} we also include $9\%$ uncertainty from jet energy scale, $7\%$ uncertainty from ISR/FSR, and $11\%$ uncertainty from the $b$ tagging efficiency.

For the ATLAS analysis~\cite{ATLAS-CONF-2011-034} we take the $t\bar t$ and the other background uncertainties from~\cite{ATLAS-CONF-2011-034} and combine the $t\bar t$ uncertainties in the $ee$, $\mu\mu$ and $e\mu$ channels as correlated and the uncertainties of the other backgrounds as partly correlated (by taking the average of the correlated and uncorrelated result due to lack of more precise information). We consider the uncertainty of the other backgrounds to be uncorrelated with that of $t\bar t$.

The systematic uncertainty of the stop signal (which is not indicated explicitly in table~\ref{tab-events} but taken into account even though its effect on the results is very small) is dominated by the theoretical uncertainty on the cross section (appendix~\ref{sec-xsec}). We also include the luminosity uncertainty quoted above and (in analogy with $t\bar t$) a $5\%$ (Tevatron) or $10\%$ (LHC) uncertainty on the acceptance and $4\%$ (Tevatron) or $10\%$ (LHC) uncertainty from the $b$ tagging efficiency where relevant.

\section{Simulation details\label{sec-simulation}}
\setcounter{equation}{0}

For both producing stop pairs and showering the stop decay products, we run {\sc Pythia}~8.145~\cite{Sjostrand:2006za,Sjostrand:2007gs} with Tune 2C~\cite{Corke:2010yf} and CTEQ 6L1 PDFs~\cite{Pumplin:2002vw} for Tevatron analyses and Tune 4C~\cite{Corke:2010yf} with the CTEQ6.5 PDFs~\cite{Tung:2006tb} (obtained via LHAPDF~\cite{Whalley:2005nh}) for LHC analyses. The intermediate step of decaying the stop NLSPs is done according to~(\ref{ME}) with our own software.\footnote{The authors will be willing to provide copies of the software, or parton-level event files produced with it, to any interested parties.} For the gravity-mediated stops that we use in the process of deriving limits from the CDF stop mass reconstruction procedure of~\cite{Aaltonen:2009sf} in section~\ref{sec-mass-rec}, the decay is performed by {\sc Pythia}, after the desired spectra are created using {\sc SuSpect}~\cite{Djouadi:2002ze}.

The output of {\sc Pythia} is further processed with our own software. Lepton identification is simulated by checking whether the leptons pass the various requirements of geometric acceptance, $p_T$, and calorimeter depositions along their trajectories, as defined in each experiment. Lepton isolation requirements are also simulated as appropriate in each case. The missing energy vector $\slash{\vv{E}}_T$ is computed based on the energy deposits in the calorimeters, as well as the full energies of muons with $|\eta| < 2$ (CDF or D0) or $|\eta| < 2.7$ (ATLAS). Jets are clustered using the {\sc FastJet}~\cite{Cacciari:2005hq} implementation of JetClu~\cite{Abe:1991ui} with cone size $R = 0.4$, overlap threshold $0.75$, seed threshold $1$~GeV and with ratcheting (for CDF analyses) or the anti-$k_T$ jet algorithm~\cite{Cacciari:2008gp} with cone size $R = 0.4$ (for ATLAS analyses). We then apply the selection requirements, including cuts on $p_T$ and $\eta$ of the various objects, $\slash E_T$, $H_T$ and other analysis-specific quantities.

For the ``tagged'' samples, we require at least one $b$ tag, simulated as follows. We determine which of the jets that satisfy the analysis conditions are $b$-jet candidates by requiring them to be within an $R = 0.4$ cone of any of the two parton-level $b$ quarks, within $\pm 60\%$ of $p_T$ of that $b$ quark, and have $|\eta| < 1.5$ (CDF) or $|\eta| < 2.5$ (ATLAS). At most two jets are allowed to become such candidates. Then we assume fixed tagging efficiency for each of them (we used $0.45$ for CDF analyses, $0.7$ for the ATLAS analysis~\cite{ATLAS-CONF-2011-034}, and $0.5$ for the ATLAS analyses~\cite{ATLAS-CONF-2011-035,Aad:2011ks}).

\begin{table}
$$\begin{array}{|c|c|}\hline
\mbox{Analysis}                                         &\,\,\, t\bar t~\mbox{scale factors}\,\,\, \\\hline
\mbox{CDF~\cite{CDF-9890}, pre-tag sample}              &   0.88,0.66,0.73     \\
\mbox{D0~\cite{Abazov:2010xm}, up to selection 1}       &   0.54                \\
\,\mbox{CDF~\cite{Aaltonen:2009sf}, $b$-tagged sample}  &   0.90,0.89,0.88  \\
\mbox{ATLAS~\cite{ATLAS-CONF-2011-034}, pre-tag sample} &\, 0.63,1.03,0.74   \\
\mbox{ATLAS~\cite{ATLAS-CONF-2011-036}}                 &   1.14             \\\hline
\end{array}$$
\caption{The ratios of the $t\bar t$ event yields predicted by our simulations to those quoted in the experimental references, for the various analyses highlighted in section~\ref{sec-results}. Scale factors are quoted for $ee$, $\mu\mu$ and $e\mu$ channels separately, where available.}
\label{tab-scale-factors}
\end{table}

\begin{table}[t]
$$\begin{array}{|c|c|c||c|c|}\hline
m_\st & m_{\tilde\chi_1^\pm} & m_{\tilde\chi_1^0} &\,\mbox{tagless}\,&\,\mbox{tagged}\,\\\hline
\, 132.5 \,&\, 105.8 \,&\, 47.6 \,& 1.2 & 1.3 \\
   135     &   125.8   &   55.1   & 1.2 & 2.1 \\
   155.8   &   105.8   &   64.9   & 1.1 & 1.1 \\
   160     &   125.8   &   58.8   & 1.1 & 1.1 \\
   168.3   &   125.8   &   71.7   & 1.1 & 1.2 \\
   179.2   &   105.8   &   61.3   & 0.9 & 1.0 \\\hline
\end{array}$$
\caption{Comparison between the results of our and CDF simulation (ratios of our and their numbers of events) for gravity-mediated stops studied in~\cite{Aaltonen:2009sf,CDF-9439,Johnson:2010zza}.}
\label{tab-CDF-DIL-stop-comparison}
\end{table}

We do not simulate certain factors such as lepton reconstruction efficiencies etc. However, we effectively take them into account to some extent by comparing our results for $t\bar t$ with those obtained by the experimentalists in their simulations, and multiplying our results for the stops by the same scale factors, as described by eq.~(\ref{eq-nstop}). The scale factors for the analyses of table~\ref{tab-events} are shown in table~\ref{tab-scale-factors}. Since even the scale factors themselves are not too far from $1$, we estimate that after applying the scale factors our simulated event yields for the stop are correct to within $\pm 10\%$. This is confirmed by table~\ref{tab-CDF-DIL-stop-comparison} where we compare our results for stops in gravity-mediated scenarios with those obtained by CDF~\cite{Aaltonen:2009sf,CDF-9439,Johnson:2010zza} (after normalizing our predictions to the stop cross sections used by CDF).\footnote{The only exception is the $135$~GeV stop in the $b$-tagged channel where our result is about twice bigger. This can probably be attributed to the fact that $m_\st - m_{\tilde\chi_1^\pm} = 9.2$~GeV, so the $b$ jets are extremely soft and the $p_T$ dependence of $b$ tagging efficiency plays an important role.}

\section{CDF stop mass reconstruction algorithm\label{sec-stop-mass-algorithm}}

We have implemented the dilepton stop mass reconstruction algorithm of~\cite{Aaltonen:2009sf,CDF-9439,Johnson:2010zza} as follows:
\begin{itemize}
\item The two highest-$E_T$ jets are assigned to the two leptons in the way that results in a smaller value of $m_{\ell_1 b_1}^2 + m_{\ell_2 b_2}^2$.
\item The possible $\vv{p}_T$ direction of each $\tilde\chi_1^0+\nu$ pair is scanned over the full $\phi$ range. At each $(\phi_1,\phi_2)$ point, the MET carried by each $\tilde\chi_1^0+\nu$ pair is then fixed by $\vv{p}_T$ conservation.
\item Each $\tilde\chi_1^0+\nu$ pair is modeled as a ``pseudoparticle'' with mass $m_0 = 75$~GeV, and the chargino mass $m_{\tilde\chi^\pm}$ is assumed to be either $105.8$ or $125.8$~GeV (the values for which the results are available in the CDF study). Then the on-shell conditions for $\tilde\chi_1^0+\nu$ and $\tilde\chi^\pm$ solve the system. This is subject to a 4-fold ambiguity since for each of the two possible pairings of $(\tilde\chi_1^0+\nu)$ with $\ell$ there are two solutions to a quadratic equation. All the solutions are kept.
\item Find the stop mass $m^{\rm rec}_\st$ that minimizes $\chi^2$ for a given $(\phi_1,\phi_2)$ point, where
\be
\chi^2 = \sum_{k=1,2}\l[\frac{\l(m^{\rm fit}_{0_k} - m_0\r)^2}{\Gamma_0^2}
+ \frac{\l(m^{\rm fit}_{{\tilde\chi_1^\pm}\,_k} - m_{\tilde\chi_1^\pm}\r)^2}{\Gamma_{\tilde\chi_1^\pm}^2}
+ \frac{\l(m^{\rm fit}_{\st_k} - m^{\rm rec}_\st\r)^2}{\Gamma_\st^2}\r]
\ee
with $\Gamma_0 = 10$~GeV, $\Gamma_{\tilde\chi_1^\pm} = 2$~GeV, $\Gamma_\st = 1.5$~GeV. The two $p_{z,0}$ (pseudoparticle $z$ momenta) and the two $m_0$ are allowed to vary in the fit in order to minimize $\chi^2$. CDF vary also the magnitudes (but not the directions) of the $4$-momenta of the leptons and jets to account for measurement errors and include the corresponding terms in $\chi^2$.
\item Average the resulting stop mass over all the choices of $(\phi_1,\phi_2)$, weighed by $e^{-\chi^2(\phi_1,\phi_2)}$.
\end{itemize}

\bibliography{stop}

\end{document}